\newcommand\Msun{\hbox{M$_\odot$}}
\newcommand\Zsun{\hbox{Z$_\odot$}}
\newcommand\kms{\hbox{km$\,$s$^{-1}$}}

\newcommand\one{\,{\sc i}}

\newcommand\tmult{\multicolumn{2}{c}}
\newcommand\hst{\textit{HST}}

\newcommand\swift{\textit{Swift}}
\newcommand\chan{\textit{Chandra}}
\newcommand\spit{\textit{Spitzer}}

\newcommand\ie{i.\,e.,}
\newcommand\eg{e.\,g.,}
\newcommand\cf{cf.}
\newcommand\etal{et~al.}

\newcommand\bb{$B_{450}$}
\newcommand\vb{$V_{606}$}
\newcommand\ib{$I_{814}$}
\newcommand\bvi{\textit{BVI}}

\newcommand\mhi{$M_\textup{\scriptsize H\one}$}
\newcommand\mmhi{M_\textup{\scriptsize H\one}}
\newcommand\mdyn{$M_\textup{\scriptsize dyn}$}

\newcommand\airac{$\alpha_\textrm{\scriptsize IRAC}$}
\newcommand\ha{H$\alpha$}

\newcommand\n{NGC~}

\newcommand{\farcs}{\mbox{\ensuremath{.\!\!^{\prime\prime}}}}

\newcommand{\ygg}{\textit{Yggdrasil}}

\documentclass[preprint2]{aastex}

\usepackage{graphicx,epstopdf,rotating,lscape}
 
\shorttitle{Stellar Populations in Compact Galaxy Groups: HCGs 16, 22, 42}
\shortauthors{Konstantopoulos \etal}

\begin{document}

\title{Stellar Populations in Compact Galaxy Groups: a Multi-Wavelength Study of  HCGs 16, 22, and 42, their Star Clusters and Dwarf Galaxies. }

\author{I.~S.~Konstantopoulos\altaffilmark{1}, 
A.~Maybhate\altaffilmark{2}, 
J.~C.~Charlton\altaffilmark{3}, 
K.~Fedotov\altaffilmark{4,5}, 
P.~R.~Durrell\altaffilmark{6}, 
J.~S.~Mulchaey\altaffilmark{7}, 
J.~English\altaffilmark{8}, 
T.~D.~Desjardins\altaffilmark{4}, 
S.~C.~Gallagher\altaffilmark{4}, 
L.~M.~Walker\altaffilmark{9}, 
K.~E.~Johnson\altaffilmark{9,10}, 
P.~Tzanavaris\altaffilmark{11,12}, 
C.~Gronwall\altaffilmark{3,13}.}

\altaffiltext{1}{Australian Astronomical Observatory, PO Box 915, North Ryde NSW 1670, Australia; iraklis@aao.gov.au. ISK is the recipient of a John Stocker Postdoctoral Fellowship from the Science and Industry Research Fund.}
\altaffiltext{2}{Space Telescope Science Institute, Baltimore, MD 21218, USA.}
\altaffiltext{3}{Department of Astronomy \& Astrophysics, The Pennsylvania State University, University Park, PA 16802, USA.}
\altaffiltext{4}{Department of Physics \& Astronomy, The University of Western Ontario, London, ON, N6A 3K7, Canada.}
\altaffiltext{5}{Herzberg Institute of Astrophysics, Victoria, BC V9E 2E7, Canada.}
\altaffiltext{6}{Department of Physics \& Astronomy, Youngstown State University, Youngstown, OH 44555.}
\altaffiltext{7}{Carnegie Observatories, Pasadena, CA  91101.}
\altaffiltext{8}{University of Manitoba, Winnipeg, MN, Canada.}
\altaffiltext{9}{University of Virginia, Charlottesville, VA.}
\altaffiltext{10}{National Radio Astronomy Observatory, Charlottesville, VA.}
\altaffiltext{11}{Laboratory for X-ray Astrophysics, NASA Goddard Space Flight Center, Greenbelt, MD 20771}
\altaffiltext{12}{Department of Physics;  Astronomy, The Johns Hopkins University, Baltimore, MD 21218}
\altaffiltext{13}{Institute for Gravitation and the Cosmos, The Pennsylvania State University, University Park, PA 16802.}

\begin{abstract}
We present a multi-wavelength analysis of three compact galaxy groups, HCGs~16,~22,~and~42, which describe a sequence in terms of gas richness, from space- (\swift, \hst, \spit) and ground-based (LCO, CTIO) imaging and spectroscopy. We study various signs of past interactions including a faint, dusty tidal feature about HCG~16A, which we tentatively age-date at $<1~$Gyr. This represents the possible detection of a tidal feature at the end of its phase of optical observability. Our HST images also resolve what were thought to be double nuclei in HCG~16C~and~D into multiple, distinct sources, likely to be star clusters. Beyond our phenomenological treatment, we focus primarily on contrasting the stellar populations across these three groups. The star clusters show a remarkable intermediate-age population in HCG~22, and identify the time at which star formation was quenched in HCG~42. We also search for dwarf galaxies at accordant redshifts. The inclusion of 33 members and 27 `associates' (possible members) radically changes group dynamical masses, which in turn may affect previous evolutionary classifications. The extended membership paints a picture of relative isolation in HCGs~16~and~22, but shows HCG~42 to be part of a larger structure, following a dichotomy expected from recent studies. We conclude that (a) star cluster populations provide an excellent metric of evolutionary state, as they can age-date the past epochs of star formation; and (b) the extended dwarf galaxy population must be considered in assessing the dynamical state of a compact group.\end{abstract}


\section{Introduction}\label{sec:intro}
The classical formulation of a Compact Galaxy Group (CG) defines an assortment of typically three or four, densely packed large galaxies \citep{hickson82,barton96}. As such, CGs represent the upper end of the surface/volume density distribution in the local universe. The recent advent of large spectroscopic studies and simulations of large-scale structure have enabled researchers to quantify the surroundings of CGs, and discover an even division between truly isolated systems and those embedded in larger groupings \citep[most notably][]{mcconnachie09cg,mendel11}. Their importance in the context of galaxy evolution therefore becomes evident when considering the possible end-states that can arise from such assortments of galaxies. Isolated groups may give rise to field ellipticals, while embedded groups might be the sites of galaxy pre-processing, where spiral galaxies deplete their gas supply and morph into lenticulars before falling into the nearest deep potential well, and eventually become the ingredients of a dry merger. 

A number of works have quantified the evolutionary state of CGs, starting from the cool gas content -- a proxy of the available reservoir for star formation. \citet{vm01} established that CGs are deficient in H\one\ gas when compared to field galaxies of like morphology, while \citet{johnson07} compared the ratio of gas-to-total mass to establish a rudimentary evolutionary sequence. In \citet{isk10} we expanded on this concept by dividing groups into dual, parallel sequences, with one track for groups where gas is contained within the member galaxies (Type~A), and another for those groups that feature an intra-group medium (IGM), be it in cold, warm, or hot gas (Type~B). The end points of the two sequences differ significantly, in that only the enhanced-IGM sequence should develop an X-ray halo, such as those seen around massive elliptical galaxies. This is an important feature, as the typically shallow CG potential well cannot build up a hot gas halo, with, perhaps, the exception of the most massive CGs. This was supported by \citet{desjardins13}, who recently found the X-ray emission in several HCGs to be concentrated mostly around individual galaxies. Collisions between galaxies and the IGM, however, can produce an X-ray halo, an effect common among groups with high velocity dispersions \citep[\eg\ in Stephan's Quintet;][]{bahcall84,sulentic95}. 

One aspect of CGs that has in the past been limited is dwarf galaxy membership. Since dense environments are likely to process dwarfs more efficiently through accretion and infall \citep{mobasher03}, the study of dwarfs is potentially a topic of particular importance to CG evolution. Only a few studies have touched upon this, such as the statistical work by \citet[][part of a study of the CG luminosity function]{hunsberger98} and the study of ultra-compact dwarf formation by \citet{darocha11}. \citet{zm98,zm00} examined the dwarf populations of several loose groups, and included HCG~42 in their study. The dwarf galaxy population of HCG~42 was further enhanced by samples observed by \citet{decarvalho99} and \citet{carrasco06}, bringing the membership to a few dozen potential dwarf members. In our previous works on HCGs~7~and~59 we included spectroscopic samples complete to $R\approx18$, but that only covers a few dwarf galaxies in each group, \eg\ the single detection in HCG~7. Our incomplete understanding of the CG luminosity function (and therefore dwarf membership) has potentially adverse effects on various measurements. While we routinely use metrics well-suited to galaxy clusters, such as the velocity dispersion, their true value when derived from three or four elements is highly uncertain \citep[as concluded by][]{mcconnachie08}. 

At the same time, the stellar populations of CGs have been the focus of numerous studies. Since star formation and galaxy evolution go hand-in-hand, much that is known about the latter has come about by studying the colors of CG galaxies \citep[\eg\ infrared colors;][]{gallagher08,bitsakis11,walker12,cluver13} and the properties of their star clusters -- be it old globular clusters \citep[GC,][]{darocha02}, or massive star clusters of young and intermediate age \citep{palma02,gallagher10,fedotov11,isk10,isk11,isk12a}. The GCs show mostly regular characteristics, with some groups hosting poor populations \citep[\n6868;][]{darocha02}, and young star clusters have been successfully used as chronometers of past dynamical events in HCG~92 \citep[Stephan's Quintet;][]{fedotov11}. While the potential is great, it comes at a great cost, as \hst\ offers the only currently suitable instrumentation for definitively distinguishing star clusters in wide extragalactic fields from foreground stars and background galaxies \citep[\eg][]{schweizer04,isk13a}. Given the propensity of CGs for interactions, intragroup stellar populations should also be common. The study by \citet{darocha05} found the intragroup light of HCG~79 to be consistent with an old population, indicative of either ancient interactions, or old stars stripped in a recent event. One evolutionary step ahead of intragroup light, in HCG~59 we found a tidal bridge connecting two galaxies, and roughly dated its emergence to within the past Gyr \citep{isk12a,white03}. 

In recent work we have pursued an understanding of galaxy evolution in CGs through multi-wavelength studies of individual groups \citep{gallagher10,isk10,isk12a} and larger samples alike \citep{hunsberger98,johnson07,gallagher08,tzanavaris10,walker10,walker12,desjardins13}. This work continues the series by contrasting the stellar populations of three CGs, HCG 16, 22, and 42, in search of evolutionary trends. We choose these groups as they encompass the full sequence of gas richness defined in \citet{johnson07} and \citet{isk10}: HCG~16 is Type~IB; HCG~22 is Type~IIA; and HCG~42 is Type~IIIA (the notation of this classification system is outlined in Section~\ref{sec:literature}). We examine the colors of young and old star clusters, which proxy their ages, to trace ongoing star formation, the recent star formation history, and past merger events. Specifically, we seek a reflection of the evolving evolutionary state of the HCG~16-22-42 sequence in the star cluster age distribution \citep[see the starburst- and interaction-dating of][]{isk09a,fedotov11}. In order to establish the star formation history diagnostic most appropriate to CGs we compare star cluster colors to broadband metrics in the UV, optical, and IR, and stacked wide-field images where we search for tidal features. Finally, we examine the dwarf galaxy populations by collating past surveys and combining with new and previously unpublished spectroscopy. We look for an assessment of the importance of including the dwarfs in characterizations of the stellar populations, and the derivation of dynamical masses.

\section{Overview of HCGs 16, 22, and 42 and Related Literature}\label{sec:literature}
Table~\ref{tab1} gives an overview of the positional, physical, morphological, photometric, and nuclear properties of the eleven galaxies in the three CGs we will be studying in this paper. The early-type galaxy content increases along the 16-22-42 sequence and the neutral gas content decreases, but other characteristics are not as neatly defined. HCG~16 hosts at least two AGN \citep{turner01} and a starburst, while star formation and nuclear activity is lower in the two other groups. Regarding the J07/K10 evolutionary type \citep{johnson07,isk10}, the roman numerals describe decreasing gas richness (Type~I is rich), while the letter discriminates between Sequence~A of groups with gas contained in individual galaxies, and Sequence~B, containing groups that show evidence of an IGM. We note that the A/B type of HCG~42 cannot be confidently constrained, as it is uncertain whether the diffuse X-ray emission is connected to the IGM, or galaxy 42A alone \citep{desjardins13}. We refer to this group as Type~IIIA throughout this paper. 

HCG~16 has been studied from several points of view. \citet{decarvalho97} updated its dwarf galaxy population by including three new members, thus increasing the velocity dispersion of the group by $\approx65~$\kms\ to 400~\kms. The merger history of this CG was investigated by \citet{decarvalho99}, who suggest that both group lenticulars, 16C and 16D, are remnants of recent mergers, as traced by double spectroscopic nuclear peaks. HCG~16 is described by \citet{belsole03} as a collapsed group at the low-luminosity end of X-ray emitters, while \citet{rich10} trace a bi-conical polar outflow in 16D with optical integral-field spectroscopy, which resembles the super-wind in M82 \citep[recently revisited by][]{desjardins13}. They also detect an intermediate-age ($\sim400~$Myr) A-star population that is rapidly rotating, consistent with the large sample of lenticular galaxies studied by \citet{atlas3d_2}. In a very recent work, \citet{vogt13} revisited the galactic winds of HCG~16. The HCG~16D wind was characterized as symmetric and shock-excited, while the outflow\newline
\clearpage\onecolumn
\begin{landscape}
\newcommand\dum{\hspace{13pt}}
%
\begin{deluxetable}{lccccc r@{.}l r@{.}l r@{$\pm$}l cr}
\tabletypesize{\footnotesize}
\tablewidth{0pt}
\tablecolumns{14}
\tablecaption{HCGs~16, 22, 42: Positional, Morphological, Photometric, and Nuclear Properties, and Measured Masses.\label{tab1}}
\tablehead{
\colhead{Identifier}   & 
\colhead{Coordinates}  & 
\colhead{Type}         & 
\colhead{$m_R$}        & 
\colhead{$v_R$}        & 
\colhead{$M*$}         & 
\tmult{\mhi}           & 
\tmult{SFR}            & 
\tmult{sSFR}           & 
\colhead{Nucleus}      & 
\colhead{References}  \\ 
                                           & 
\colhead{(J2000)}                          & 
                                           & 
\colhead{(mag)}                            & 
\colhead{(\kms)}                           & 
\multicolumn{3}{c}{($\times10^{9}~$\Msun)} & 
\tmult{(\Msun~yr$^{-1}$)}                  & 
\tmult{($\times10^{-10}~\textup{yr}^{-1}$)}& 
                                           & 
}
\startdata
\sidehead{\textbf{HCG~16}}
A: \n0835         & 02~09~24.6~$-$10~08~09  & SBab & 12.30  & 4073  & 2.65  & 1&17        &  5&$37\pm0.62$\tablenotemark{*} &\dum$3.68\,$&$\,0.57$ & AGN    & [1], [1], [7], [13]  \\
B: \n0833         & 02~09~20.8~$-$10~07~59  & Sab  & 12.65  & 3864  & 1.03  & 0&79        &  0&$33\pm0.03$\tablenotemark{*} &  $0.43\,$& $\,0.07$  & AGN    & [1], [1], [7], [13]  \\
C: \n0838         & 02~09~38.5~$-$10~08~48  & Im   & 12.82  & 3851  & 1.19  & 3&02        & 14&$38\pm1.83$                  &  $21.19\,$&$\,4.10$  & SB     & [2], [1], [1], [13]  \\
D: \n0839         & 02~09~42.9~$-$10~11~03  & Im   & 13.86  & 3874  & 1.03  & $>4$&47     & 17&$06\pm2.31$                  &  $30.10\,$&$\,5.80$  & AGN    & [3], [5], [6], [13]  \\
\sidehead{\textbf{HCG~22}}
A: \n1199         & 03~03~38.4~$-$15~36~48  & E2   & 10.95  & 2570  & 2.09  &\tmult{$-$}  & 0&$24\pm0.02$                   &  $0.21\,$& $\,0.03$  & A      & [2], [6], [9], [14]  \\
B: \n1190         & 03~03~26.1~$-$15~39~43  & Sa   & 13.64  & 2618  & 0.14  &\tmult{$-$}  & 0&$03\pm0.01$                   &  $0.38\,$& $\,0.07$  & A      & [4], [6], [3], [14]  \\
C: \n1189         & 03~03~24.5~$-$15~37~24  & SBcd & 13.19  & 2544  & 0.15  & $<1$&35     & 0&$46\pm0.04$                   &  $5.64\,$& $\,0.95$  & E      & [3], [6], [10], [14] \\
\sidehead{\textbf{HCG~42}}
A: \n3091         & 10~00~14.3~$-$19~38~13  & E3   & 10.31  & 3964  & 4.73  & 4&35        & 0&$44\pm0.04$\tablenotemark{*}  &  $0.17\,$& $\,0.03$  & AGN    & [3], [6], [9], [14]  \\
B: \n3096         & 10~00~33.1~$-$19~39~43  & SB0  & 13.03  & 4228  & 0.80  &\tmult{$-$}  & 0&$10\pm0.03$                   &  $0.23\,$& $\,0.07$  & A      & [4], [6], [3], [14]  \\
C: MCG~-03-26-006 & 10~00~10.3~$-$19~37~19  & E2   & 12.86  & 4005  & 0.86  &\tmult{$-$}  & 0&$09\pm0.02$                   &  $0.18\,$& $\,0.04$  & A      & [4], [6], [11], [14] \\
D: PGC~028926     & 10~00~13.0~$-$19~40~23  & E2   & 14.73  & 4042  & 0.13  &\tmult{$-$}  & 0&$01\pm0.01$                   &  $0.15\,$& $\,0.07$  & A      & [3], [6], [12], [14] \\
\enddata
%
\tablecomments{%
  Morphological types from \citet{hickson89}.
  Stellar masses, star formation rates (SFR) and 
  specific SFRs (sSFR) are drawn from \citet{tzanavaris10}, 
  corrected for a known $M_*$ overestimation factor of 7.4 
  (Tzanavaris, private communication).  
  \mhi\ values from \citet[][individual galaxies in HCG~16]{vm01};
  \citet[][HCG~22]{price00};
  and \citet{huchtmeier94} for the entirety of HCG~42, contained
  within the quoted HCG~42A beam. 
  Nuclear classifications of `A' and `E' stand for absorption- 
  and emission-line dominated spectra. 
  %
  References are given in sequences representing positions, 
  magnitudes, radial velocities, and nuclear classification: 
     [1]:~\citet{sdss}; [2]:~\citet{chandracat}; 
     [3]:~\citet{decarvalho97}; [4]:~\citet{2mass}; 
     [5]:~\citet{hopcat}; [6]:~\citet{hickson89};
     [7]:~\citet{ribeiro96}; [8]:~\citet{hyperleda2}; 
     [9]:~\citet{huchtmeier94}; [10]:~\citet{monnier03}; 
     [11]:~\citet{hickson92}; [12]:~\citet{carrasco06};
     [13]:~\citet{turner01}; [14]:~Tzanavaris~\etal, in preparation. 
  Asterisks ($^*$) denote SFR values that are potentially 
  contaminated by AGN. 
  }
\end{deluxetable}
\end{landscape}
\twocolumn

\noindent in 16C was found to be asymmetric due to the interaction with the H\one\ envelope into which the wind is advancing, and the excitation mechanism was attributed to a blend of photoinisation and slow shocks. 

The globular cluster population of HCG~22A was studied with ground based optical imaging by \citet{darocha02}, who estimate a specific frequency ({number of clusters per unit brightness}) of $S_N = 3.6\pm1.8$, consistent with its morphological type of E3. The globular clusters trace a bimodal color distribution with peaks at $(B-R)_0=1.13, 1.42$~mag, a common occurrence in early-type galaxies \citep[\eg][]{brodiestrader06}. \citet{darocha11} discovered a population of 16 ultra-compact dwarf (UCD) galaxies in HCG~22, and used them to suggest two channels of UCD formation: old, metal-poor star clusters, and stripped dwarf galaxies with higher metallicities and mixed (or young) stellar populations. 

As described above, HCG~42 was one of the poor groups studied by \citet{zm98,zm00} as part of a dwarf galaxy survey. Combined with \citet{decarvalho97} and \citet{carrasco06}, these works have identified dozens of galaxies related to the group. We will be following up on this aspect with particular interest over the following sections. 

\section{Observations}\label{sec:observations}
The work presented in this paper is based on new and archival data from ground- and space-based observatories, which we summarize in Table~\ref{tab:obs_im}. We use optical imaging from the Las Campanas Obseratory (LCO) DuPont telescope (Direct CCD Camera and Wide Field Imaging CCD Camera) in the B and R bands,  taken at the same time as the images presented in Konstantopoulos et al. (2010). We therefore refer the reader to that paper for details on the observational setup, as well as the data acquisition and reduction. In the case of HCG~16, the LCO images are complemented by Sloan Digital Sky Survey \citep[SDSS;][]{sdss} imaging in the $ugriz$ filter-set. 

Multi-wavelength imaging comes from \chan, \swift, \textit{Hubble} (\hst), and \spit\ space telescopes. {Archival data are drawn from the \chan\ X-ray Center and the NASA/IPAC Infrared Science Archive for post-basic calibration \spit\ imaging in the four IRAC bands. We make use of the same} Swift UV/Optical Telescope (UVOT) dataset presented in \citet[][we refer the reader there for details]{tzanavaris10}, which includes images in the \emph{w2}, \emph{m2}, and \emph{w1} filters \citep[central wavelengths of 1928, 2246, and 2600~\AA;][]{poole08}. \hst\ images from the WFPC2 and ACS/WFC cameras were reduced on-the-fly using the \textit{Mikulski Archive for Space Telescopes}. WFPC2 data were then processed with \texttt{MultiDrizzle} to register each chip separately, before correcting the photometry for charge transfer inefficiencies using the prescription of \citet{dolphin00cte}. \hst\ coverage consisted of multiple pointings, as listed on Table~\ref{tab:obs_im}. 

We also make use of new spectroscopy from CTIO-Hydra and previously unpublished spectra from the DuPont telescope (Multifiber Spectrograph and 2D-FRUTTI). The Hydra observations were taken with a combination of the KPGL2 grating ($R\sim4400$) and GG385 blocking filter, tuned to a central wavelength of 5800~\AA, and are described in \citet{isk10}. The Magellan spectra were taken on the same observing run as the data used for \citet{zm98}, so we refer readers there for details of the acquisition and reduction. In brief, the wavelength coverage extends between $3500-6500~$\AA, at a resolution of $\sim5-6~$\AA\ (3~\AA\,px$^{-1}$ dispersion).

\begin{table*}[htbp]
\begin{center}

\newcommand\ind{\hspace{5pt}}
\renewcommand\sidehead[1]{%
 \noalign{\vskip 1.5ex}%
 \multicolumn{6}{l}{\hspace{-6pt}#1} \\%
 \noalign{\vskip .5ex}%
}

\caption{Summary of Imaging Observations}\label{tab:obs_im}
\begin{tabular}{lccccc} 
\tableline
\tableline
Target							& 
Instrument						& 
Filter							& 
Date							& 
$t_\textup{\scriptsize exp}$	& 
Program ID						\\
		& 
		& 
		& 
		& 
(sec)	& 
		\\
\tableline
\sidehead{\textbf{\hst}}
HCG~16	& WFPC2		& F435W			& 2007-07-17	& 1900	& 10787	\\
\ldots	& \ldots	& F606W			& 2007-07-17	& 1900	& \ldots\\
\ldots	& \ldots	& F814W			& 2007-07-23	& 1900	& \ldots\\
HCG~22	& \ldots	& F435W			& 2007-09-21	& 1900	& \ldots\\
\ldots	& \ldots	& F606W			& 2007-09-21	& 1900	& \ldots\\
\ldots	& \ldots	& F814W			& 2007-09-21	& 1900	& \ldots\\
HCG~42	& \ldots	& F435W			& 2007-11-13	& 4200	& \ldots\\
\ldots	& \ldots	& F606W			& 2007-11-13	& 4200	& \ldots\\
\ldots	& \ldots	& F814W			& 2007-11-13	& 4200	& \ldots\\
\ldots	& ACS-WFC	& F435W			& 2007-12-04	& 1710	& \ldots\\
\ldots	& \ldots	& F606W			& 2007-12-06	& 1230	& \ldots\\
\ldots	& \ldots	& F814W			& 2007-12-08	& 1080	& \ldots\\
\sidehead{\textbf{\swift}}
HCG~16	& UVOT		& UVW2			& 2007-02-24	& 4652	& --\tablenotemark{1}\\
\ldots	& \ldots	& UVM2			& 2007-02-24	& 3894	& --	\\
\ldots	& \ldots	& UVW1			& 2007-02-24	& 2596	& --	\\
HCG~22	& \ldots	& UVW2			& 2007-03-17	& 3650	& --	\\
\ldots	& \ldots	& UVM2			& 2007-03-17	& 3214	& --	\\
\ldots	& \ldots	& UVW1			& 2007-03-17	& 2524	& --	\\
HCG~42	& \ldots	& UVW2			& 2007-02-01	& 3326	& --	\\
\ldots	& \ldots	& UVM2			& 2007-02-01	& 3027	& --	\\
\ldots	& \ldots	& UVW1			& 2007-02-01	& 2017	& --	\\
\sidehead{\textbf{LCO-100"}}
HCG~16	& CCD		& JB			& 2007-10-05	& 180	& --	\\
\ldots	& \ldots	& KC-R			& 2007-10-03	& 120	& --	\\
HCG~22	& \ldots	& JB			& 2007-10-06	& 120	& --	\\
\ldots	& \ldots	& KC-R			& 2007-10-04	& 120	& --	\\
HCG~42	& WFCCD		& B				& 2008-05-08	& 110	& --	\\
\ldots	& \ldots	& R				& 2008-05-06	& 600	& --	\\
\sidehead{\spit}
HCG~16	& IRAC		& $3.6-8.0~\mu$m & 2005-01-17	& 27	& 3596	\\
HCG~22	& \ldots	& $3.6-8.0~\mu$m & 2005-01-17	& 27	& \ldots\\
HCG~42	& \ldots	& $3.6-8.0~\mu$m & 2004-12-17	& 27	& \ldots\\
\tableline
\tablenotetext{1}{The \swift\ observations were taken as a Team Project, 
	and are not associated with a program ID.}
\end{tabular}
\end{center}
\end{table*}

\subsection{Star cluster selection}\label{sec:select}
The star cluster analysis that will follow in Section~\ref{sec:clusters} is based, for the most part, on WFPC2 images. The exception is the cluster population of HCG~42A, which makes use of ACS data. In this case, we follow the methodology presented in \citet{isk10,isk12a} to select clusters. In brief, we selected star clusters with \texttt{IRAF-DAOfind} and filter according to a metric of central concentration that is based on point-spread function photometry (employing a custom-made function), before applying a brightness cut at $M_V < -9~$mag.

\begin{figure}[!h] 
	\begin{center}

		\includegraphics[width=0.49\textwidth]{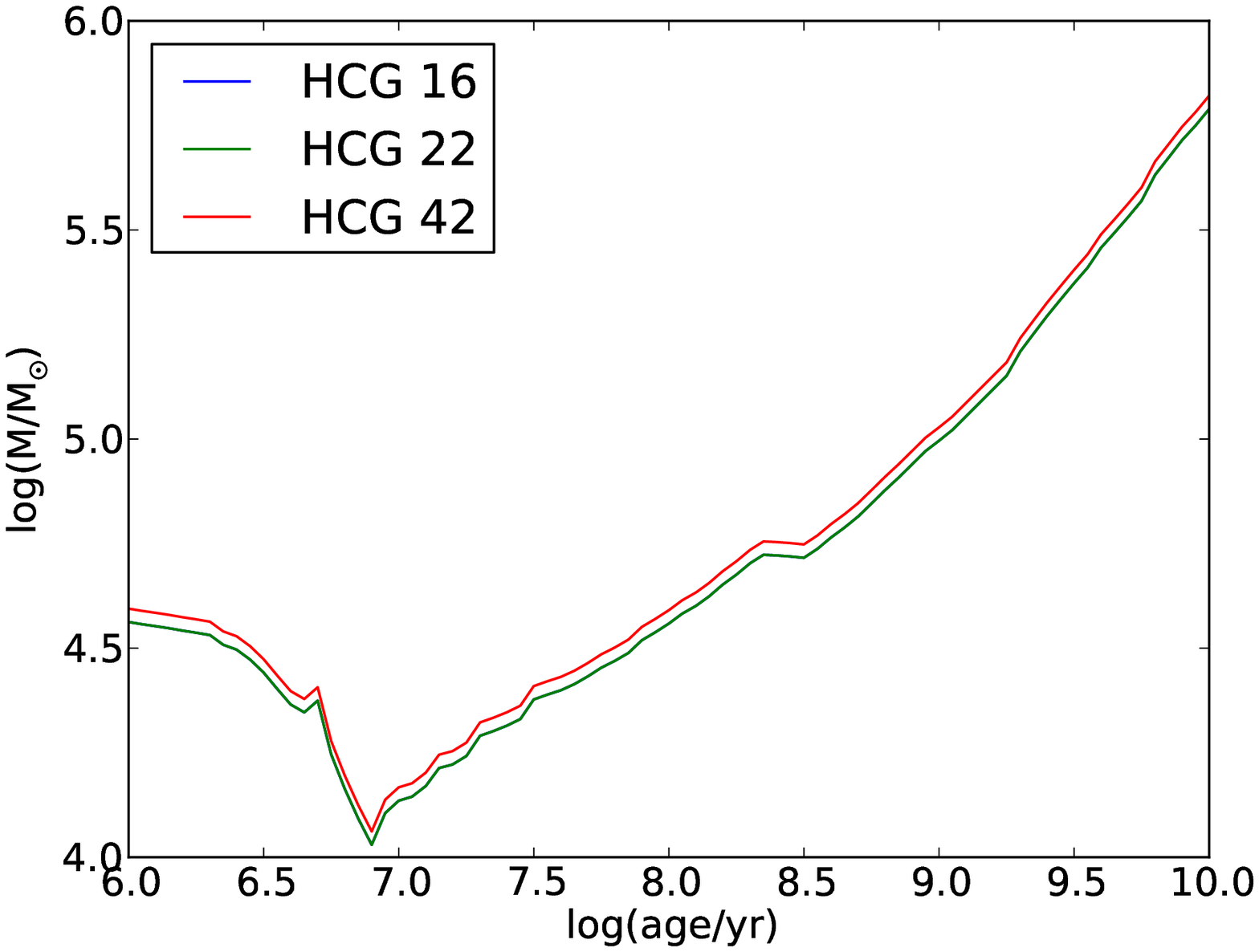}
		\caption{The conversion of $M_V=-9~$mag into mass
			over the full evolutionary course of a model simple 
			stellar population \citep{marigo08}. As a star cluster 
			ages it progressively loses stellar material (an effect
			of stellar evolution), which results in an ever decreasing 
			mass-to-light ratio. The mass required for a star 
			cluster to be detected therefore increases over time. 
			In this case, we expect to detect many clusters of 
			moderate mass in their first Gyr of evolution and
			only those globular clusters (age~$>1~$Gyr) more massive 
			than $10^5$~\Msun. The lines for HCGs~16 and 22 overlap. 
		}\label{fig:agemass}

	\end{center}
\end{figure}

In the case of WFPC2, the lower resolution does not allow for such filtering. Instead, we performed a more thorough selection process rather than filtering a long list. First we divided each image by the square root of its median to create a frame with uniform noise. Such a detection image was created for each individual chip, each pointing, and each filter. We then performed an \texttt{IRAF-DAOfind} search and cross-correlated the resulting detection lists, promoting only sources that appeared in two or more lists to the final source catalogue. Photometry was then performed in each original image. Unfortunately, it was not possible to further filter through WFPC2 source photometry, so we did so through visual inspection of all sources with $M_V < -9~$mag. This eliminated the vast majority of background galaxies, as their surface brightness profiles differ largely from those of star clusters. While it is not trivial to quantify the expected degree of contamination by background galaxies, we note that spirals are expected to span the entire sequence of star clusters colors (from a few Myr to globular cluster colors), depending on their star formation rate and history, and dust content. This would preclude regions in color space from being devoid of data points. As will be shown in Section~\ref{sec:stellarpops}, we do observe such voids, which suggests the degree of contamination from background galaxies is low. Not all stars were removed, however, as those are largely indistinguishable from clusters at the distances studied and with the WFPC2 image scale. We therefore expect some stellar contamination at the red end of star cluster color space. 

\begin{figure*}[phtb] 
	\begin{center}

		\includegraphics[width=0.45\textwidth]{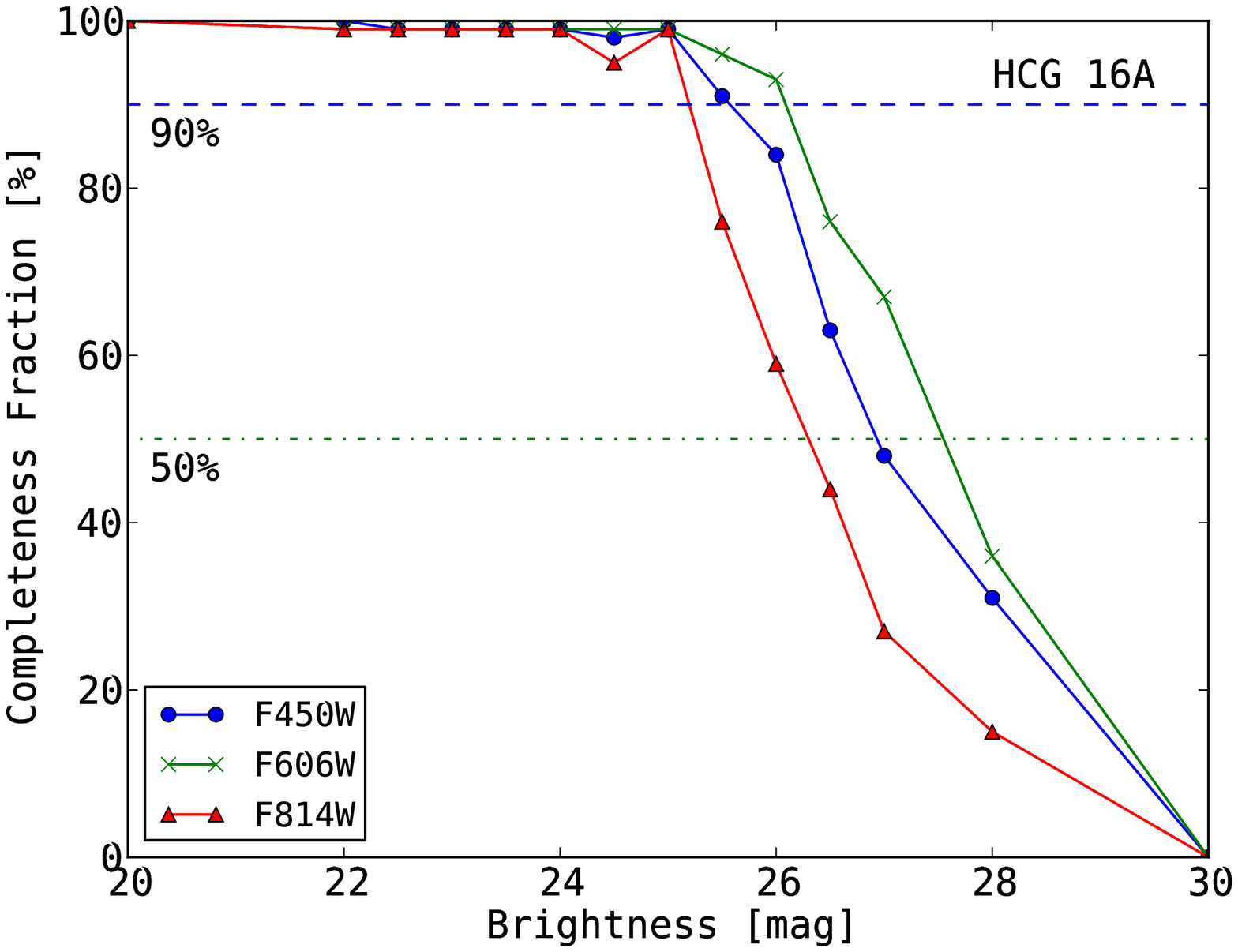}
		\includegraphics[width=0.45\textwidth]{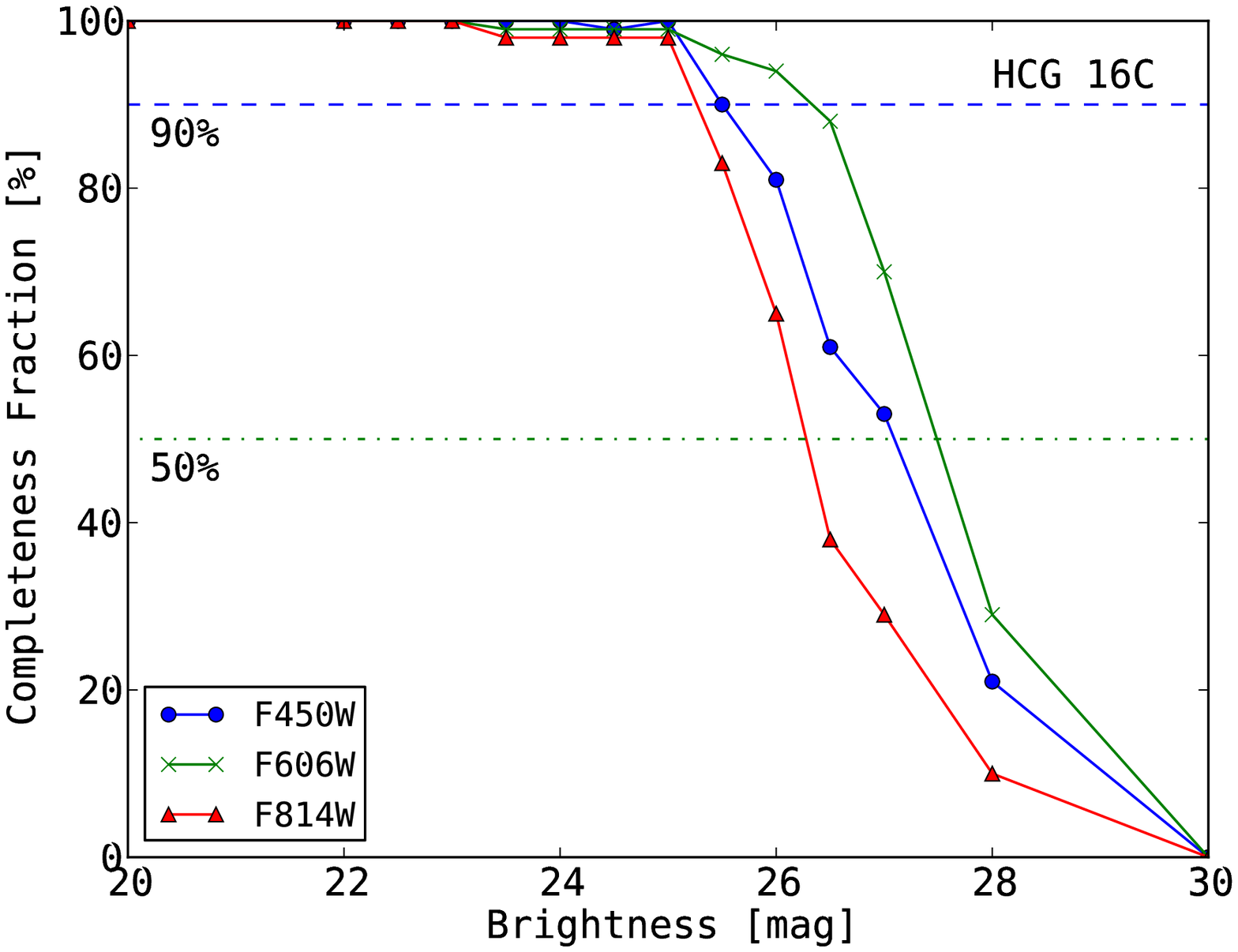}\\
		\includegraphics[width=0.45\textwidth]{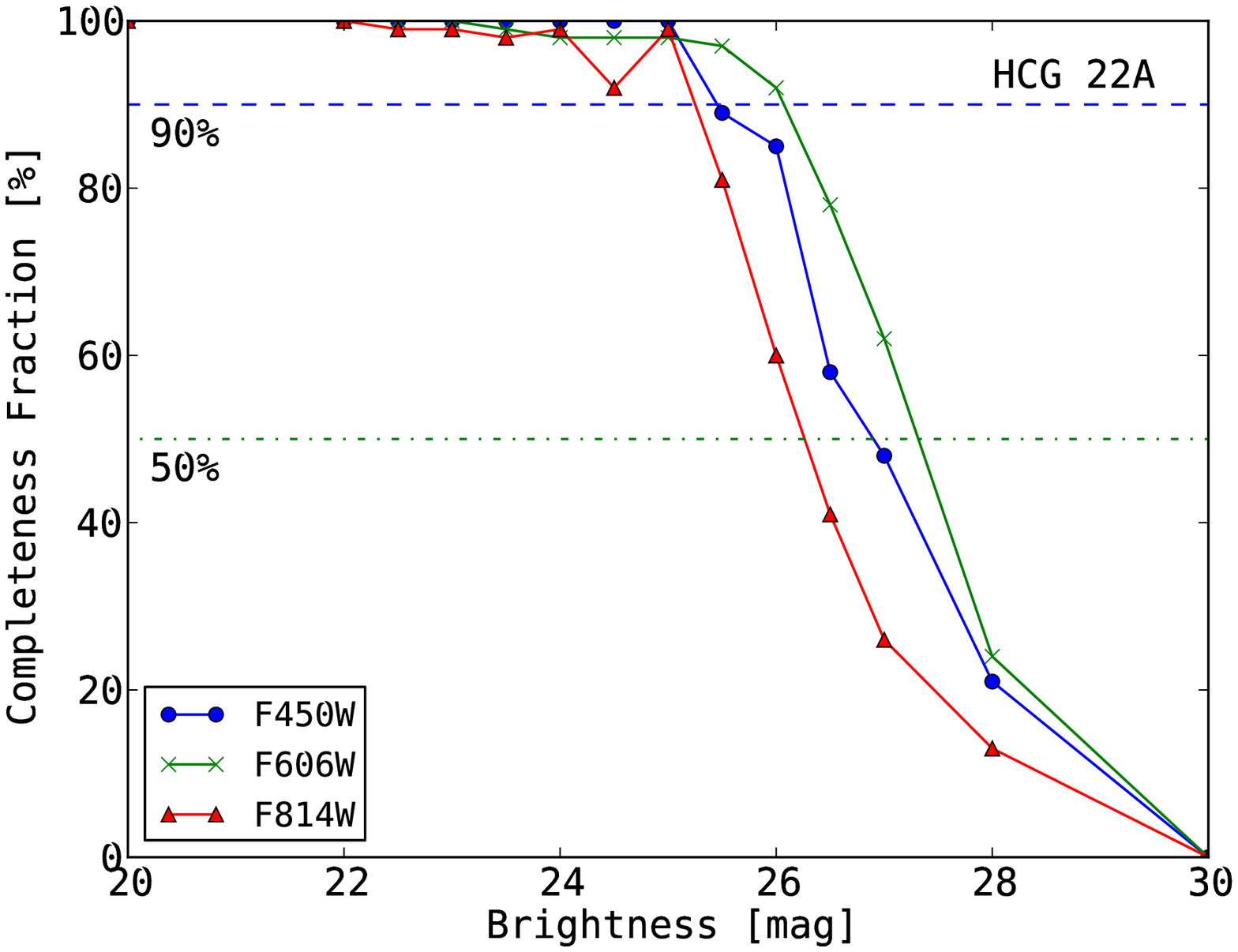}
		\includegraphics[width=0.45\textwidth]{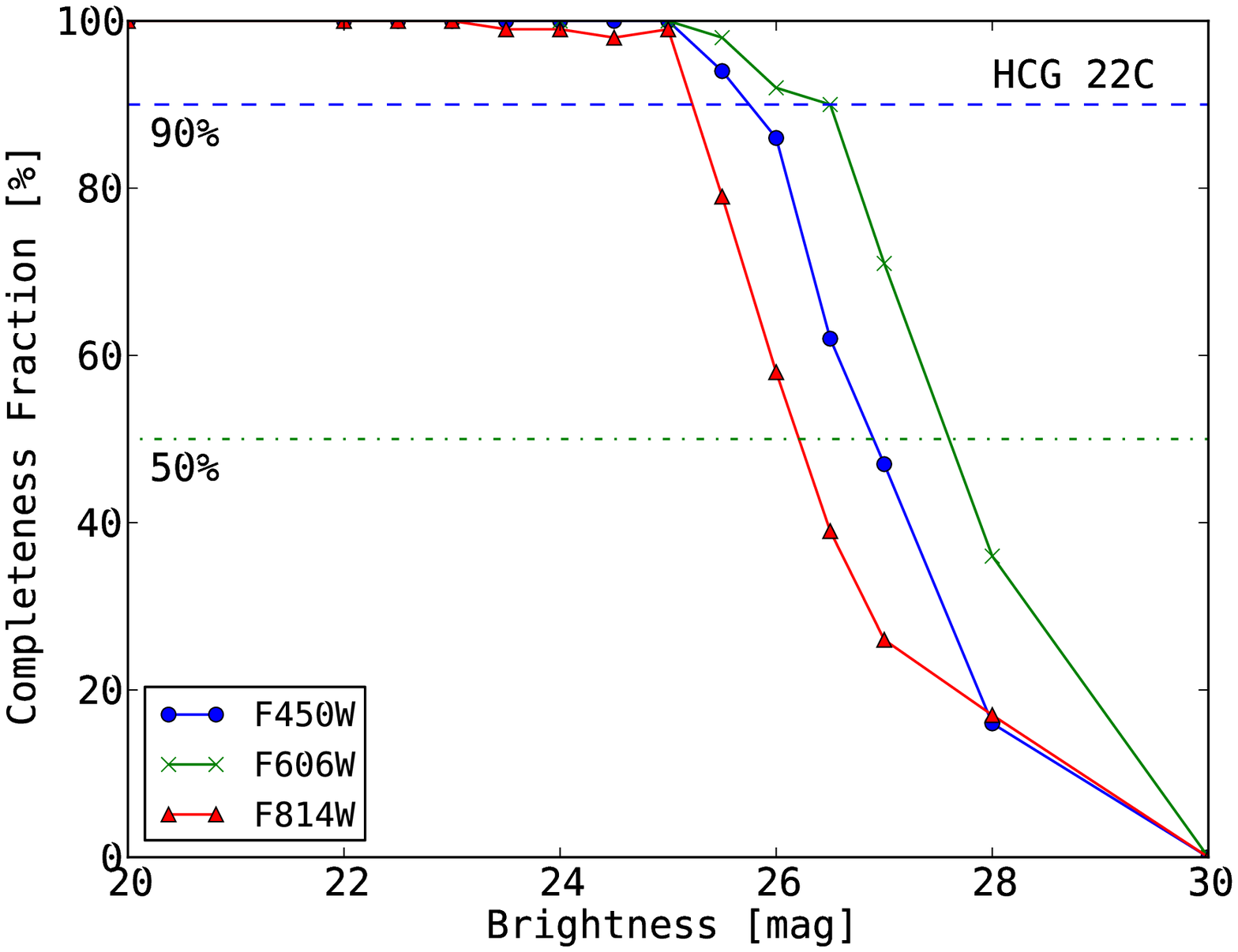}\\
		\includegraphics[width=0.45\textwidth]{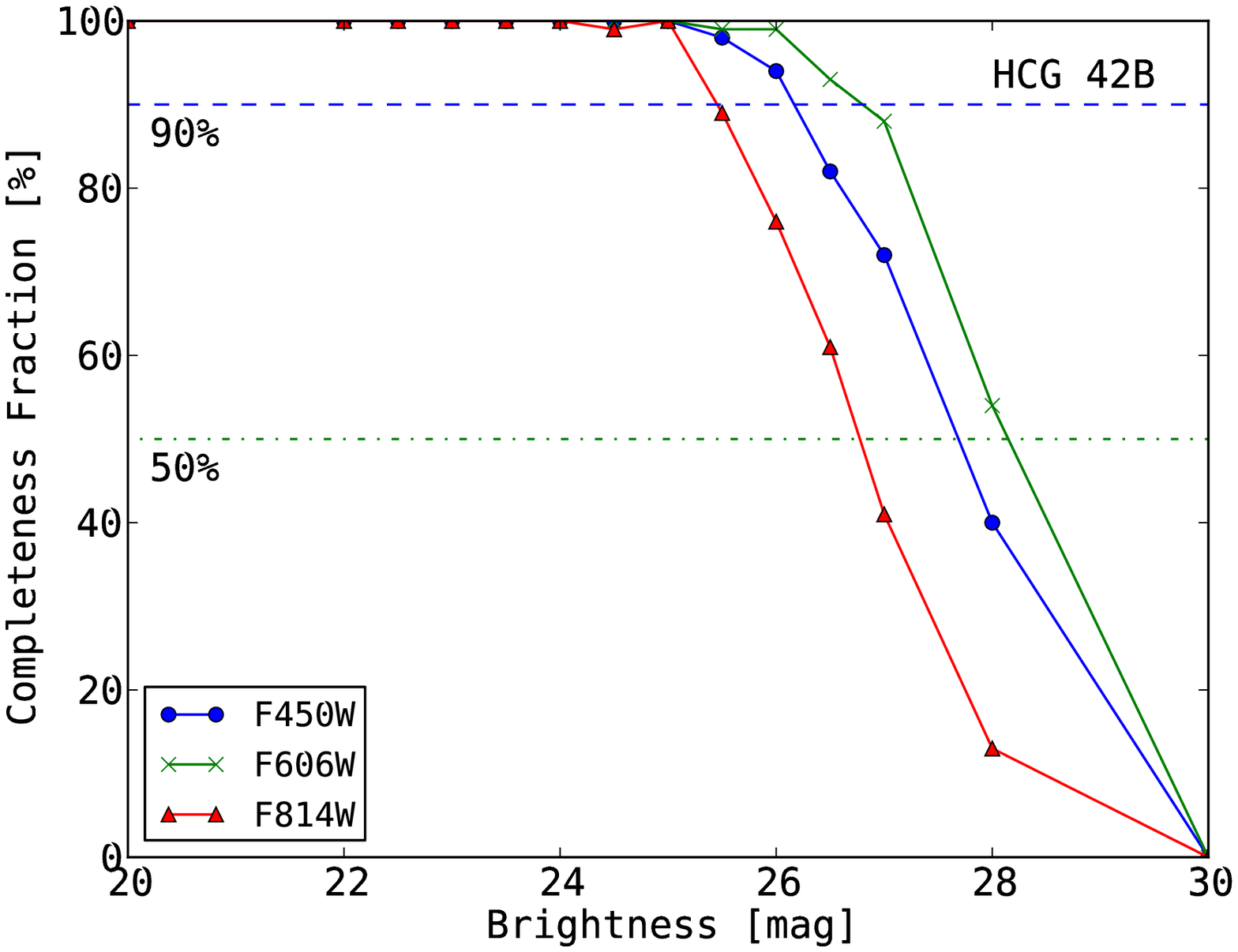}
		\includegraphics[width=0.45\textwidth]{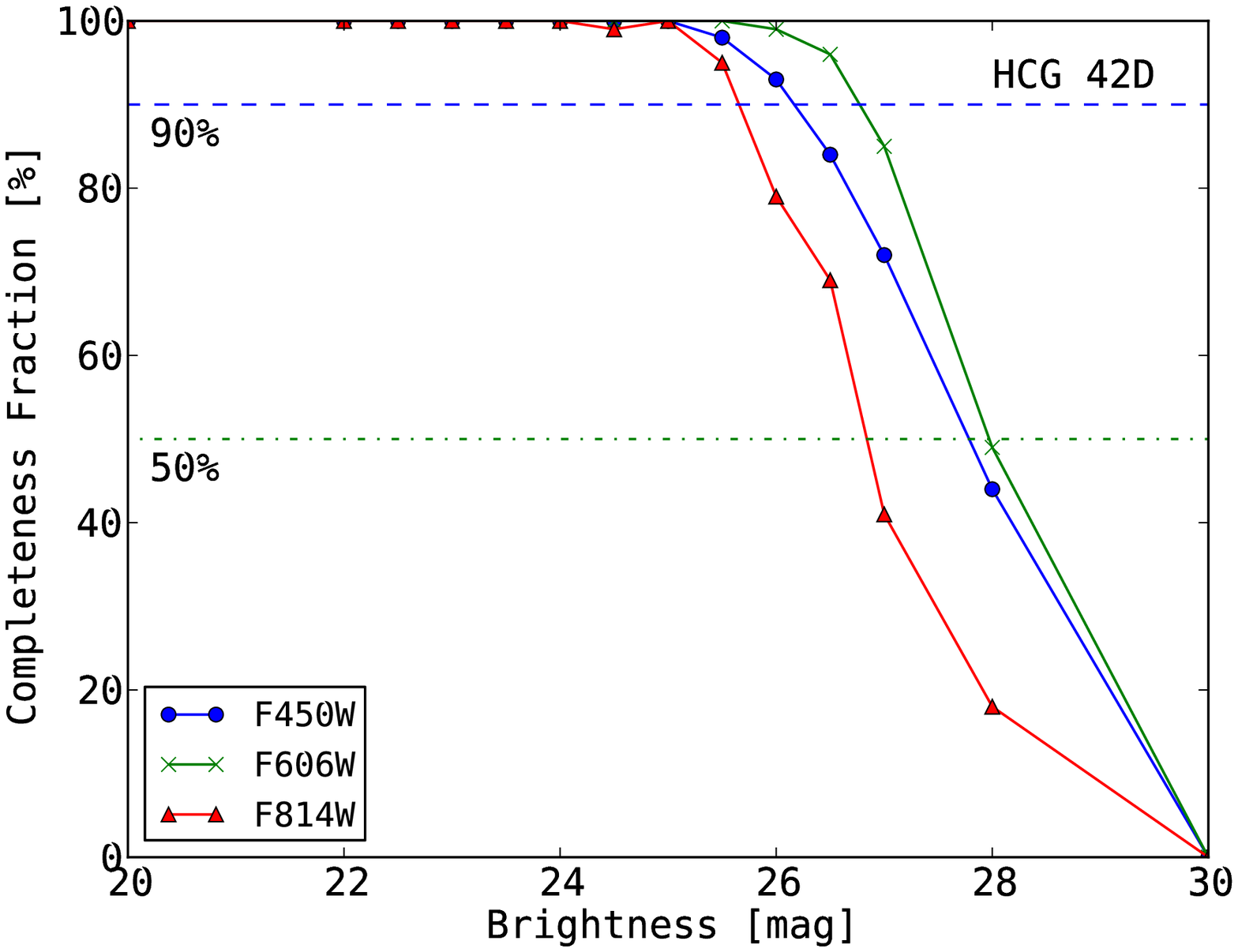}
		%
		\caption{Completeness tests for star cluster detections in
			HCGs~16, 22, 42. We test two galaxies per group to cover
			different environments, namely early- vs late-type
			galaxies. \ib\ is always the limiting filter, and defines
			the 90\% and 50\% completeness fractions as roughly 25, 
			26~mag in HCGs~16, 22, and 25.5, 26.5~mag in HCG~42. 
		}\label{fig:completeness}

	\end{center}
\end{figure*}

The 50\% and 90\% completeness fractions for our WFPC2 star cluster selection were estimated by generating artificial sources with \texttt{MKSynth} \citep[part of the \texttt{BAOLAB} suite;][]{ishape} and testing their recovery through the detection method described above. We generated a $10\times10$ grid of artificial sources and inserted them into the images of two galaxies per group, in order to cover varying morphologies. The grid covered both galaxy and background, as we wish to assess the observability of star clusters throughout these CGs. Our estimates therefore present a best-case scenario for spiral and irregular systems, where crowding and variable extinction will further complicate detection. However, the detection limits should be taken at face value in the case of of early-type galaxies, as their smooth profiles do not inhibit detection. Figure~\ref{fig:completeness} plots the curves, from which we derive [90\%, 50\%] completeness fractions of roughly [25, 26]~mag in HCGs~16, 22, and [25.5, 26.5]~mag in HCG~42. These are all fainter than the brightness cut applied at $M_V=-9~$mag, therefore we treat our star cluster catalogues as luminosity-limited. It is important in studies of star cluster populations to understand the detection limit not only in terms of brightness, but also mass. Since the mass-to-light ratio of a simple stellar population (SSP) evolves with time, the mass required for a star cluster to be detected will change with its age. Figure~\ref{fig:agemass} investigates this relation by translating the limiting brightness of a \citet{marigo08} model SSP into the corresponding mass. We estimate completeness over the past $[1, 10]~$Gyr to masses of $[1, 5]\times10^5~$\Msun. 

\section{Phenomenology}\label{sec:phenomenology}
\begin{figure*}[phtb] 
	\begin{center}

		\includegraphics[width=270pt,angle=270]{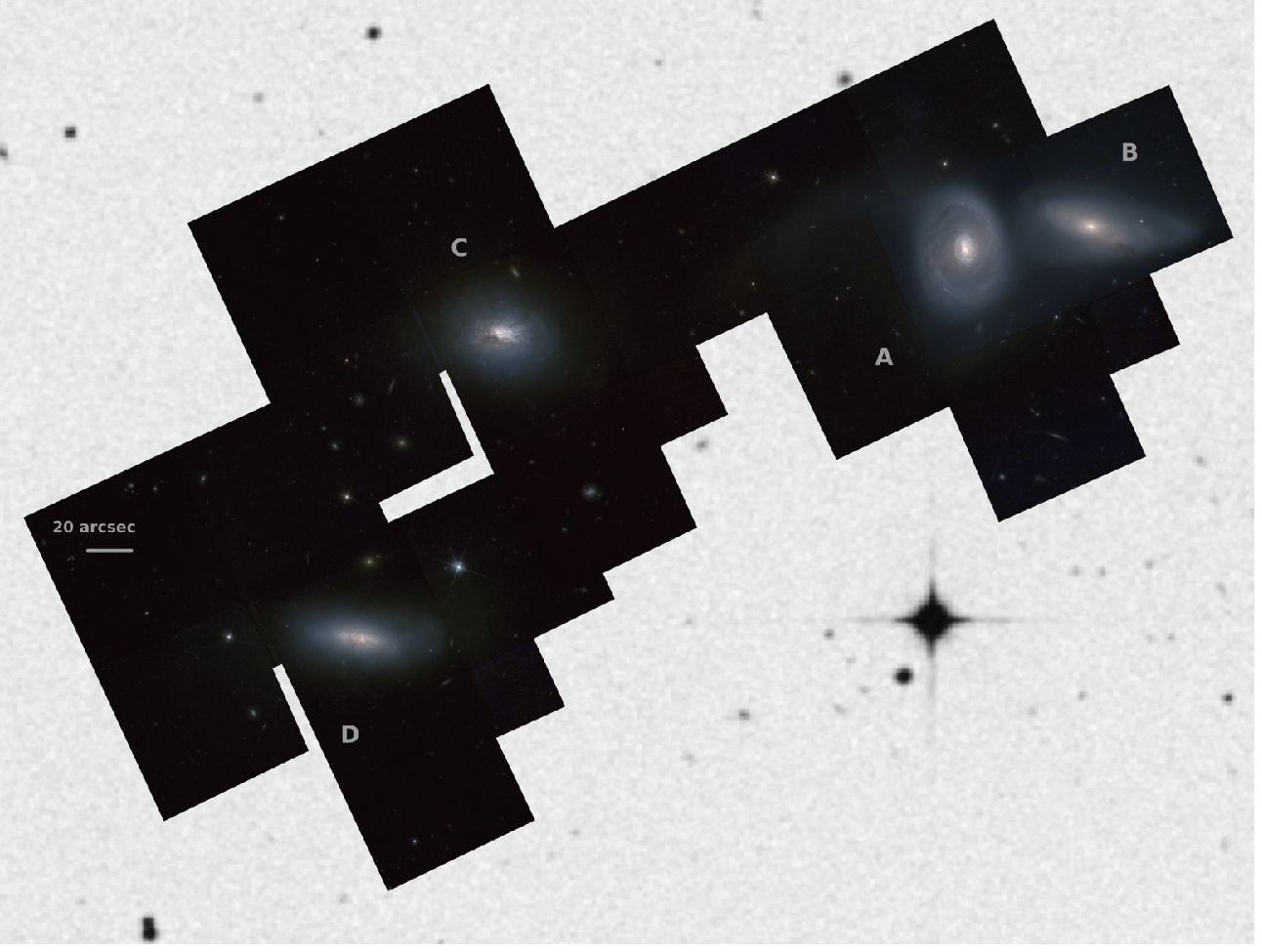}\\
		\includegraphics[width=315pt,angle=0]{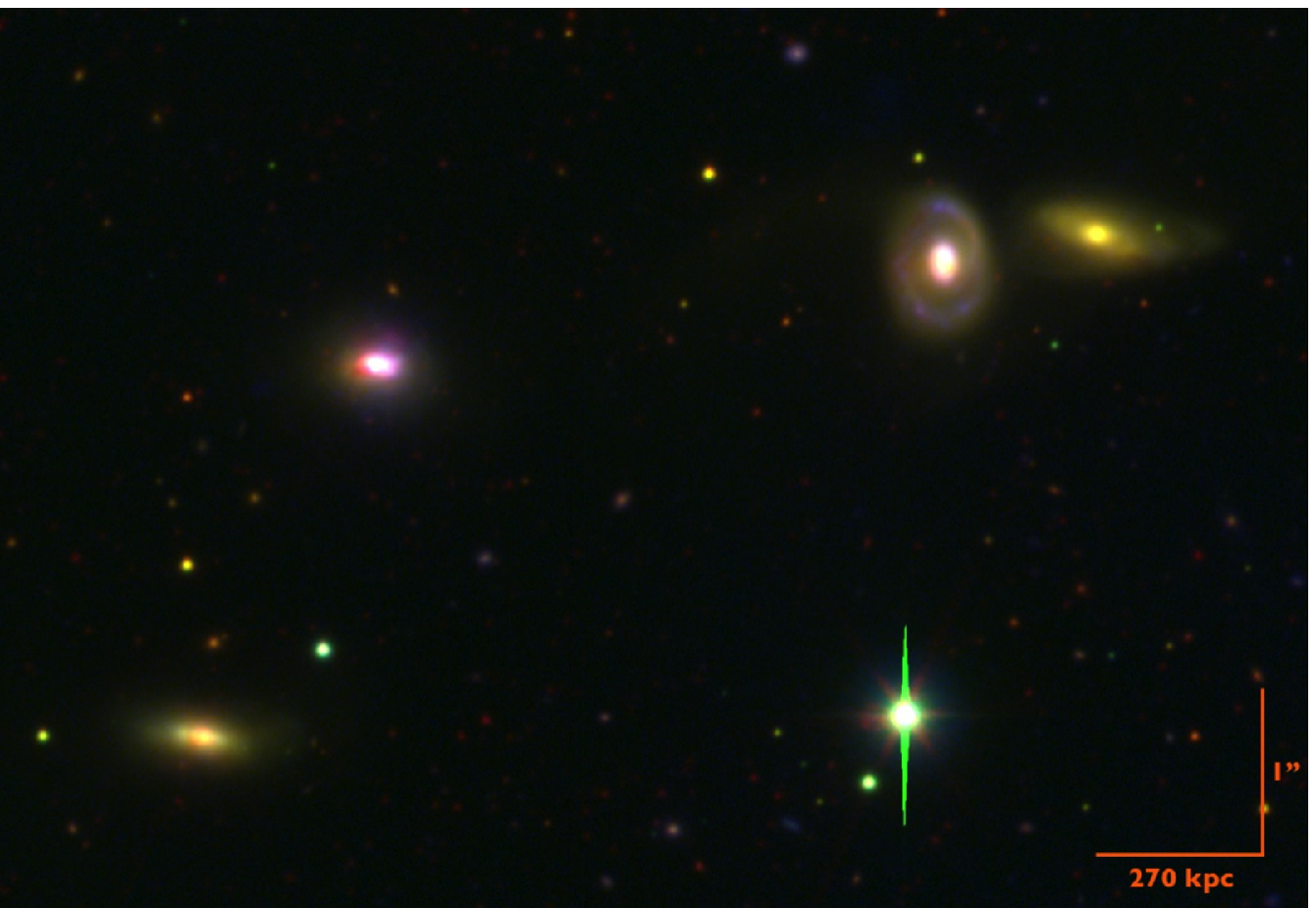}
		\caption{\textbf{Top}: \hst-\bvi\ imaging of HCG~16, 
			framed in a DSS grayscale background. 
			The \textbf{bottom} image makes use of  \swift-UV, 
			LCO~R-band, and \spit~$3.6~\mu$m frames. Star 
			formation registers blue and cyan colors, while 
			a yellow or red appearance reveals older stellar 
			populations. 
			The star-forming and interacting nature of HCG~16 
			is evident in this image: galaxies~C and D appear 
			patchy and dusty in the \hst\ image, while galaxy~B 
			displays an asymmetric profile, despite its 
			early-type classification. Galaxy~A shows significant 
			star formation activity in its slender spiral arms. 
		}\label{fig:finder16}

	\end{center}
\end{figure*}

\begin{figure*}[phtb]
	\begin{center}

		\includegraphics[width=273pt,angle=0]{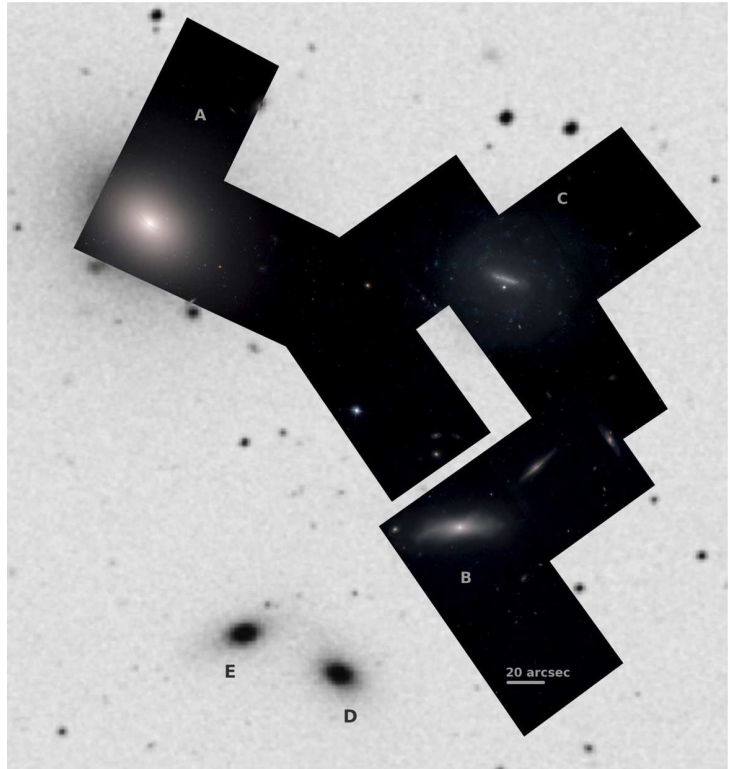}
		\includegraphics[width=312pt,angle=0]{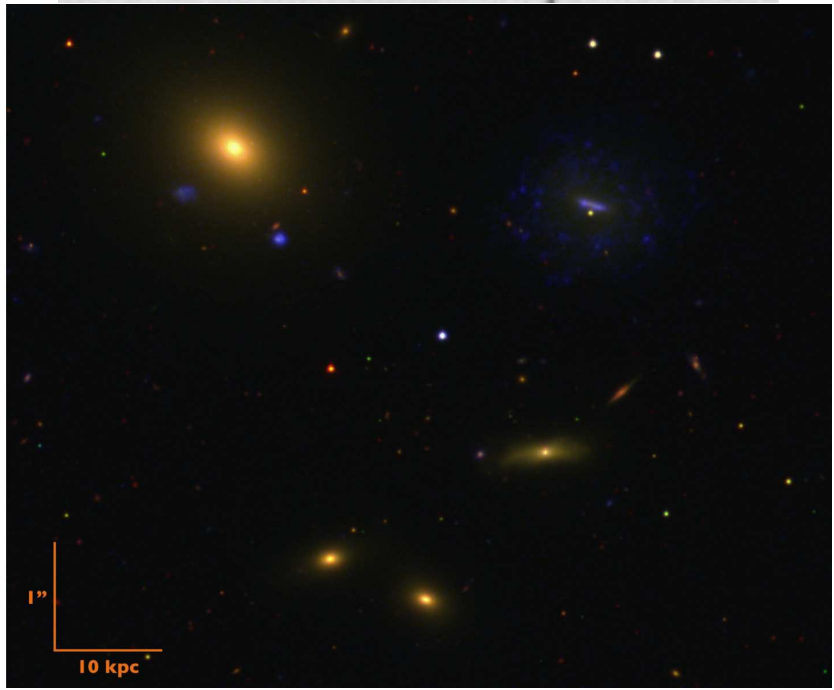}
		\caption{
			As in Figure~\ref{fig:finder16}, but for HCG~22. This 
			group presents a more quiescent nature, with ongoing
			star formation only in the long network of spiral arms
			around galaxy~C, stemming from a bright bar. Galaxies
			A and B appear yellow in the multi-wavelength image, 
			however both show indication of mergers or interactions 
			in the past (see Figure~\ref{fig:detail22}): 22A displays 
			a thick dust lane, while 22B is highly irregular. 
			HCG~22E and D to the south-east of C are not group 
			members, but a background pair. 
		}\label{fig:finder22}

	\end{center}
\end{figure*}

\begin{figure*}[htbp]
	\begin{center}

		\includegraphics[width=300pt,angle=270]{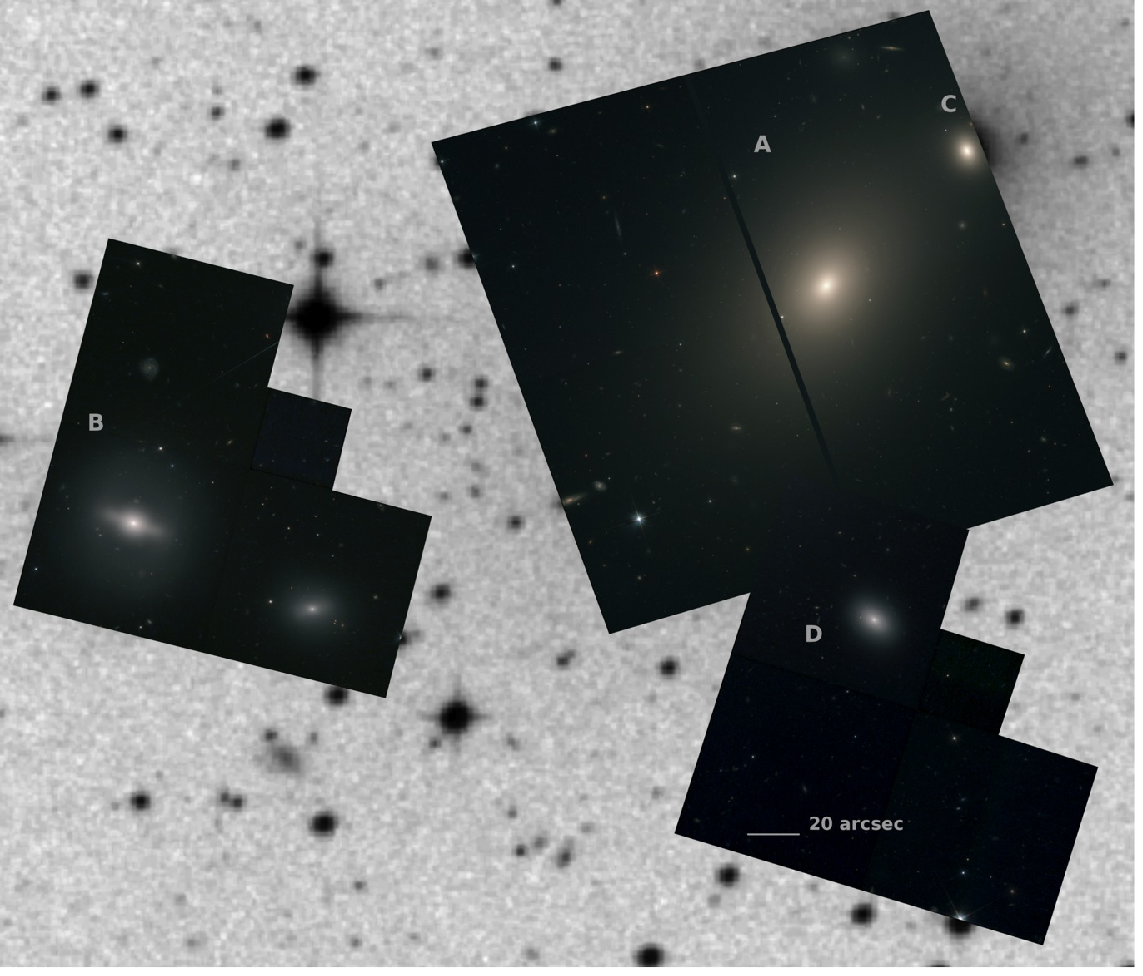}
		\includegraphics[width=330pt,angle=0]{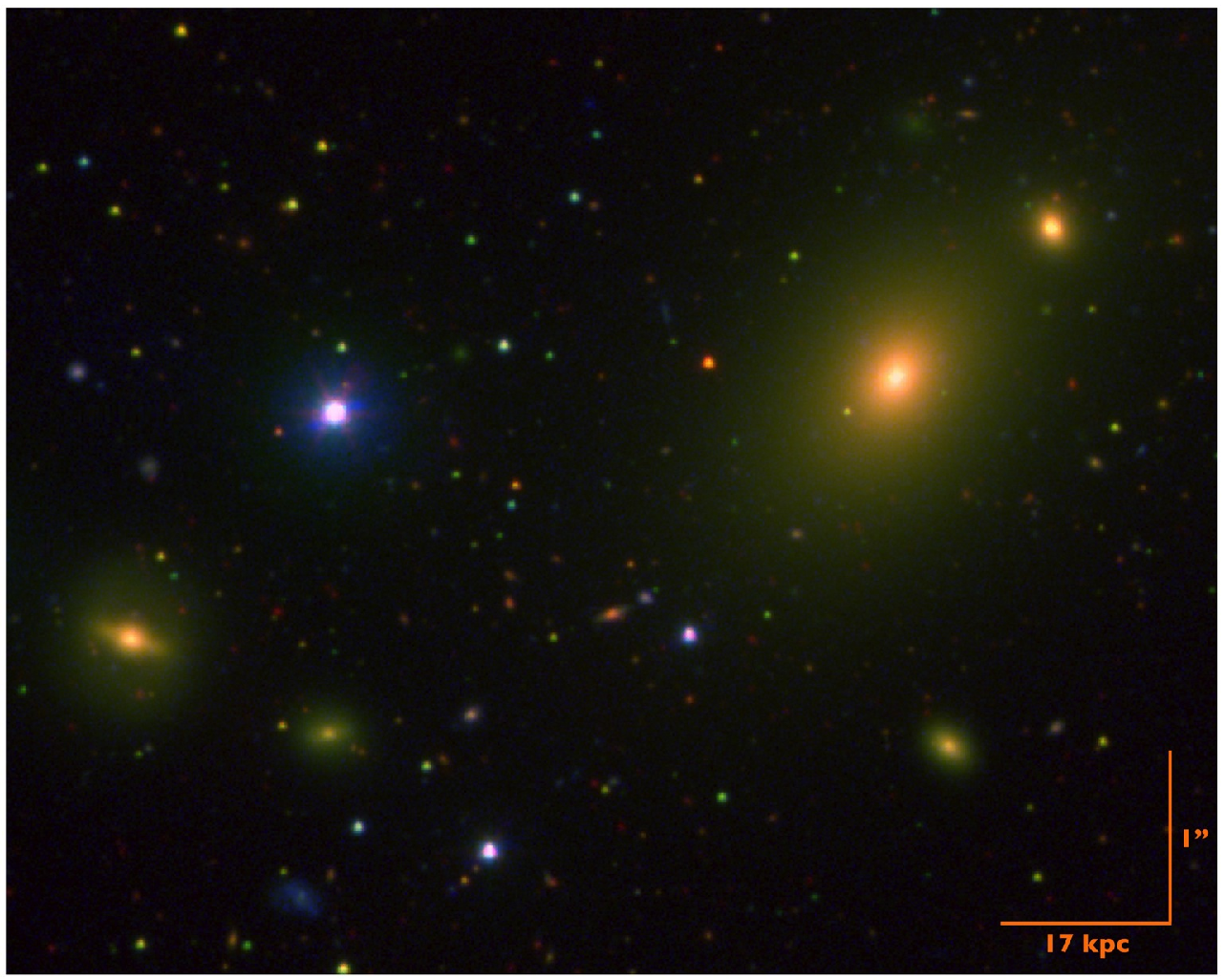}
		\caption{
			As in Figure~\ref{fig:finder16}, but for HCG~42, which
			is populated solely by quiescent galaxies. 42B shows a 
			`boxy' bulge/disk morphology, while the smaller system 
			to its west is a dwarf member (see Figure~\ref{fig:pspace}).
			We note no other peculiarities in this system. 
		}\label{fig:finder42}

	\end{center}
\end{figure*}

Figures~\ref{fig:finder16}--\ref{fig:finder42} show \hst\ imaging (BVI) and color-composites of \swift\ (UV, blue), LCO (R band, green), and \spit\ ($3.6~\mu$m, red) frames, which we use to study the large-scale properties of the groups and individual galaxies. Blue emission indicates star formation over the past $\approx100~$Myr, while older populations shine in red and yellow (photospheres shine in V-band continuum and $3.6~\mu$m). H$\alpha$ emission is covered in the R-band, so ongoing star formation will appear purple of pink. Visual inspection shows all galaxies to be consistent with their published morphological types. We note a lopsided appearance in HCG~16B, highlighted by the isophotes of the high-contrast image of Figure~\ref{fig:tail}. The `patchy' dust distribution of HCG~16C is seen both as reddening in the optical, and a red/white appearance in the multi-wavelength imaging. We also note the strong bar of HCG~22C and the low surface brightness of its spiral arms, the only part of the image that registers significant UV flux. Finally, we note a `boxy' bulge-disk appearance in HCG~42B, both in the optical image, and in stellar photospheric emission (R band, $3.6~\mu$m). The following sections will visit small-scale features in HCGs~16, 22, and 42.


\subsection{Central sources in HCG~16C and D}
\begin{figure*}[tb]
	\begin{center}
		\includegraphics[width=\linewidth, angle=0]{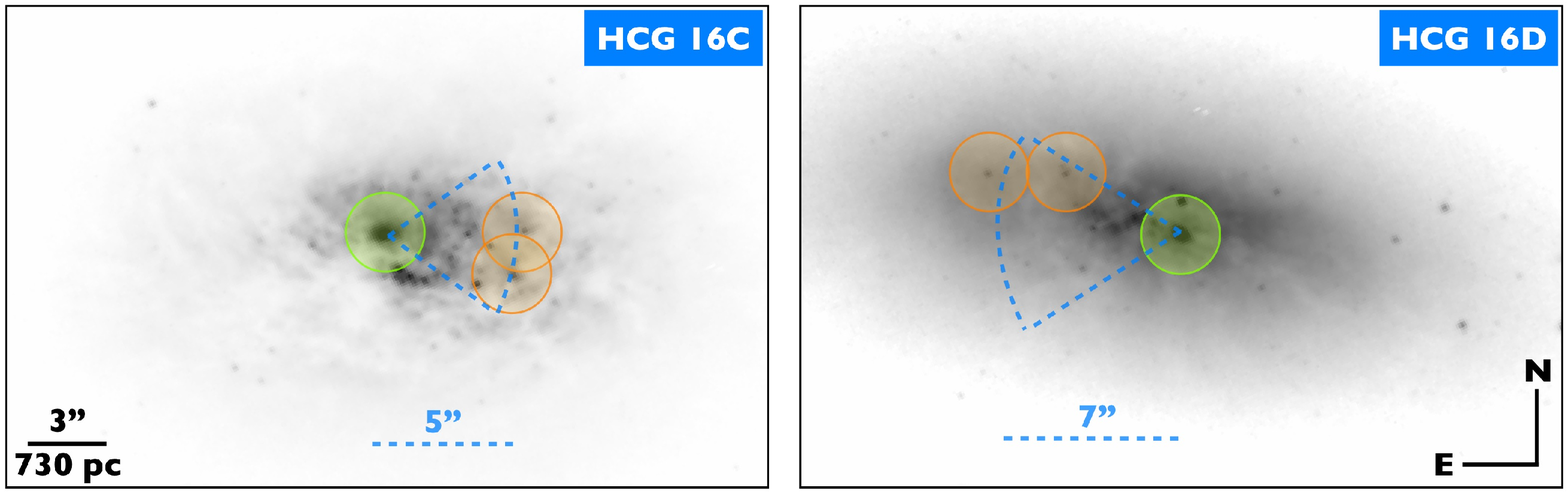}
		\caption{
			\hst-WFPC2 close-up of HCG~16C and 16D in the F814W 
			band, showing a plethora of star-forming regions. 
			The various markings are meant to relate this 
			imaging to the double nuclei reported by \citet{decarvalho99}, 
			with circles representing the size of the spectroscopic 
			extraction apertures. We mark the nuclei in green 
			circles, and then draw dashed blue arcs to indicate the 
			suggested separation between the two nuclei 
			\citep[in the direction reported by][]{decarvalho99}. 
			Orange circles mark star clusters that intersect this 
			arc and are therefore viable candidates for the detections
			flagged as second nuclei. 
			The high spatial resolution of this imaging thus argues 
			against the interpretation of the discussed previous work. 
			At the proposed distances from the respective nuclei of 
			1.2~and 1.7~kpc, the proposed second nuclei of HCG~16C~and~D 
			would give rise to an Antenn\ae- like, major merger 
			appearance, whereas the two galaxies are more reminiscent 
			of post-interaction systems, such as M82. 
			}
		\label{fig:h16cd}

	\end{center}
\end{figure*}

Taking advantage of the high resolution of the \hst\ images, we identify various bright, compact sources in the central regions of HCG~16C~and~D, as shown in Figure~\ref{fig:h16cd}. The exponential surface brightness profile of these sources, as well as their colors, are consistent with a star clusters, as will be discussed in Section~\ref{sec:clusters}. In order to understand the structure of these two galaxies we consider the WFPC2 imaging in the context of the double nuclei proposed for both HCG~16C~and~D by \citet{decarvalho99}. This report was of the spectroscopic discovery of second nuclei situated 5\arcsec\ west and 7\arcsec\ east of the main nuclei of galaxies 16C~and~D respectively. These distances are represented by the dashed blue arcs of Figure~\ref{fig:h16cd}. The arcs are intersected by star clusters, which we mark in orange circles of diameter 3\arcsec, the width of the \citeauthor{decarvalho99} spectroscopic aperture. At the $\approx30~$Mpc distance to HCG~22 the angular separations between the suggested double nuclei correspond to physical distances of $1.2~$kpc and $1.7~$kpc, suggestive of a major merger morphology akin to the Antenn\ae. This is not the case in these galaxies, which instead have morphologies reminiscent of post-interaction systems, such as M82. In addition, the $1\farcs5$ median seeing in the \citeauthor{decarvalho99} observations would have blended the light of many sources in these crowded star-forming inner regions, giving the semblance of a single, bright source, rather than a collection of star clusters. Our high resolution imaging do not support the double nucleus scenario of \citet{decarvalho99} in HCG~16C~and~D. Finally, it is likely that the AGN-like emission line ratios discovered by this previous work represent contamination from shocks in the galactic wind, which we know to be present from the analysis of \citet{vogt13}. 

\subsection{A tidal feature about HCG~16A}\label{sec:lsb}
\begin{figure}[tbhp]
	\begin{center}

		\includegraphics[width=\linewidth, angle=0]{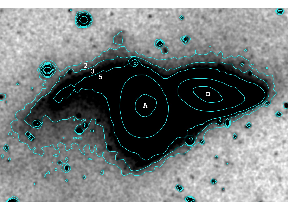}\\
		\caption{Smoothed, co-added LCO images in the B, R
			bands, presented at high contrast. North is to 
			the top, East to the left, and the field covers 
			a region of $\approx60\times40~$kpc (or
			$\approx250\arcsec\times170\arcsec$). Contours 
			correspond to the R band image, and numbers count 
			the sigma level with respect to the background. 
			A tidal feature to the East of galaxy~A is detected 
			at the $3-5\,\sigma$ level, while the isophotes 
			of HCG~16B are distorted. 
		}\label{fig:tail}

	\end{center}
\end{figure}

We make use of our deep, wide-field imaging from LCO to search for low surface brightness features in the three groups. We detect a faint tail ($3-5\,\sigma$ level with respect to the R band background) off the eastern side of 16A, shown in Figure~\ref{fig:tail}. The disturbed morphology of 16B and the ring of star formation about the center of 16A suggest that the two are involved in a tidal interaction. Given that the feature is co-spatial with the large H\one\ envelope around \mbox{16A/16B} \citep{vm01,borthakur10}, we assume that it contains gas. Tidal features are known to be detectable for no more than $\sim0.5~$Gyr in the optical, after which H\one\ is a more appropriate tracer \citep{mullan11}. Therefore, the proposed interaction was likely a recent one. 

\begin{figure}[tbhp]
	\begin{center}

		\includegraphics[width=\linewidth, angle=0]{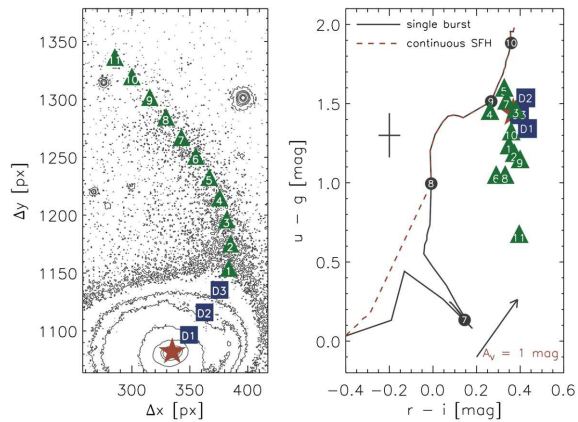}
		\caption{SDSS $ugriz$ photometry of the HCG~16A tidal 
			tail. The \textbf{left panel} shows a contour plot 
			of the $r$-band SDSS image, with the location of 
			numbered photometric apertures marked in boxes 
			(disk), triangles (tail) and a star for the nucleus. 
			The orientation has been adapted for illustrative 
			purposes (east to the top, north to the right, 
			or a clockwise $90^\circ$ rotation with respect 
			to Figure~\ref{fig:finder16}). {Given the marginal 
			detection, any deductions are tentative and need 
			to be followed up with deeper imaging}.
			The \textbf{right panel} shows the $u-g$ versus 
			$r-i$ colors of these apertures, plotted against 
			\ygg\ stellar population models representing a 
			single burst population (solid), and a continuous 
			star formation history (dashed). Numbers on the 
			track denote each age dex, while the crosshair 
			indicates typical photometric errors. When compared 
			to simple stellar populations, most apertures 
			appear to exhibit a relatively high dust attenuation, 
			as demonstrated by the reddening vector, an unusual 
			trait for tidal debris. We place the age of the 
			underlying population between $\sim100~$Myr and 
			$\sim1~$Gyr with no discernible age-space trends. 
			The clump at the end of the tail is an exception, as 
			its color suggests the presence of red supergiants 
			and hence ongoing star formation. Deeper imaging would 
			better constrain the age of the feature and hence 
			the interaction that created it. 
		}\label{fig:tail_phot}

	\end{center}
\end{figure}

We perform photometry on the tail, in order to assess its age and provenance -- did it originate as stripped gas in which stars subsequently formed \citep[\eg][]{hibbard94,knierman03,werk08}, or was it a stream of old stars to begin with \citep[\eg][]{gallagher_parker}? We use archival images from SDSS, as they cover a broader optical baseline than our LCO imaging and provide high quality flat fielding. Figure~\ref{fig:tail_phot} shows the placement of square apertures on the left (orientation has been altered for illustrative purposes) and optical colors on the right, plotted on top of a \citet{marigo08} SSP model track. We measure very low fluxes in these apertures ($3-5\,\sigma$ above background), hence the following results require confirmation from deeper imaging\footnote{Toward that end, we recently obtained deep imaging as part of a Canada-France Hawaii Telescope campaign, which will be presented in future work.}.
The measured colors are mostly representative of an aging population, between 100~Myr and 1~Gyr. We should note that it is difficult to distinguish between an aging population with a contribution from red supergiants, and a highly extinguished coeval population, based solely on \bvi\ observations. Box 11, at the tip of the tail, is an exception, as its color is highly suggestive of red supergiant stars (when accounting for the reddening vector). The clumps at the extremities of tidal tails are often found to outshine their main-body counterparts and host  more prolonged bursts of star formation \citep[\eg][]{mullan11}. They are also the formation sites of short-lived tidal dwarf galaxies in dynamical models \citep{bournaud09} and H\one\ observations \citep[\eg][]{hibbard96}, although optical spectroscopy often paints a picture of tail clumps as chaotic, unbound systems \citep[\eg][]{trancho12}. 

The star formation history of this debris feature, as deduced tentatively from Figure~\ref{fig:tail_phot}, characterizes it as an elusive event: an aging tidal tail with little ongoing star formation. Past optical studies of tidal debris have favored bright, blue, clumpy star-forming features as they are more readily observable. As a result, optical tails are never observed to contain stellar populations older than a few hundred Myr \citep{trancho07a,trancho12,bastian09antennae,fedotov11}. After that stage, they are usually only observed in radio wavelengths \citep[\eg][]{rogues,koribalski05}, with ages between $0.5-1~$Gyr \citep[inferred mostly from dynamical modeling, \eg][]{yun94}. The tail in HCG~16A is faint and smooth, and also appears to potentially be quite reddened at the $A_V\lesssim2~$mag level -- \cf\ $A_V<0.5~$mag across the \citet{mullan11} sample. The roughly inferred age of the stellar population in the 16A tail is typical of H\one\ tails, rather than those routinely studied in the optical. Combining all this information, we suggest that this might be a rare case of a tidal tail reaching the end of its optically detectable phase. Deeper observations are required to confirm this detection and our interpretation, while another possibility is that the feature is instead a stellar stream stripped from the early-type galaxy HCG~16B. 

\subsection{Other optical traits and infrared spectral energy distributions}\label{sec:seds}
\begin{figure*}[tbhp]
	\begin{center}

		\includegraphics[width=\linewidth, angle=0]{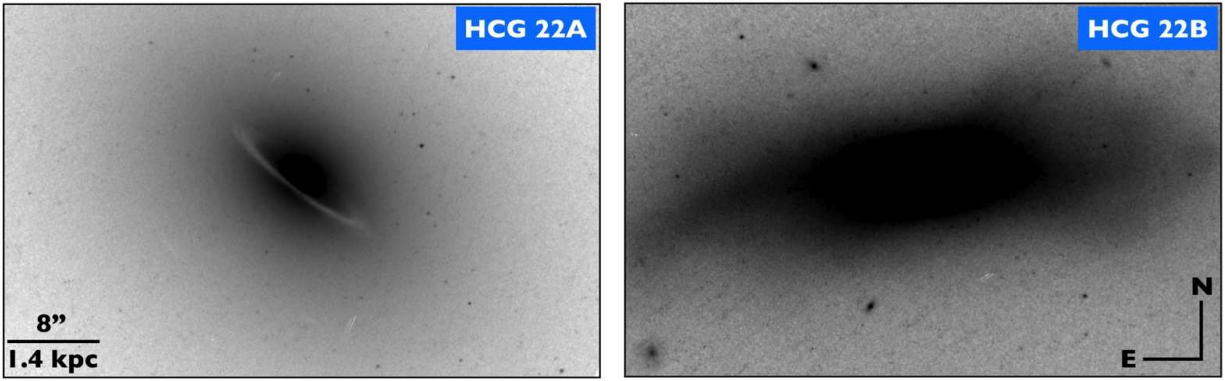}
		\caption{\textbf{Left}:~A prominent dust lane can be seen 
			in this \hst-F450W image of the inner $\approx2~$kpc 
			of HCG~22A, possibly the signature of a recent merger. 
			North is to the top and East to the left of the 
			image, as is the image on the \textbf{right}, where
			the contrast of the F814W frame has been set 
			to demonstrate the various disturbances in the optical 
			morphology of HCG~22B. We interpret this as the result 
			of a recent merger (following a series of passages).
		}\label{fig:detail22}

	\end{center}
\end{figure*}

All galaxies in HCG~22 display morphological peculiarities. We note a thick equatorial dust lane in galaxy~22A, observed in the past by \citet{sparks85}, and various low surface brightness features in 22B, probably indicative of recent mergers or infall events (Figure~\ref{fig:detail22}). The group as a whole does not show much evidence for star formation away from the faint, extended spiral arms of 22C, which consists of a small, bright bar structure, surrounded by very faint, but orderly and symmetric spiral arms. 

HCG~42 is populated exclusively by early-type galaxies with no deviations from regular morphologies, except perhaps the seemingly `boxy' (bulge/disk) light profile of 42B. The group is dominated by 42A, which features a high luminosity and stellar mass (see Table~\ref{tab1}). 

\begin{sidewaysfigure*}[p]
	\begin{center}

		\includegraphics[width=0.329\textwidth,angle=0]{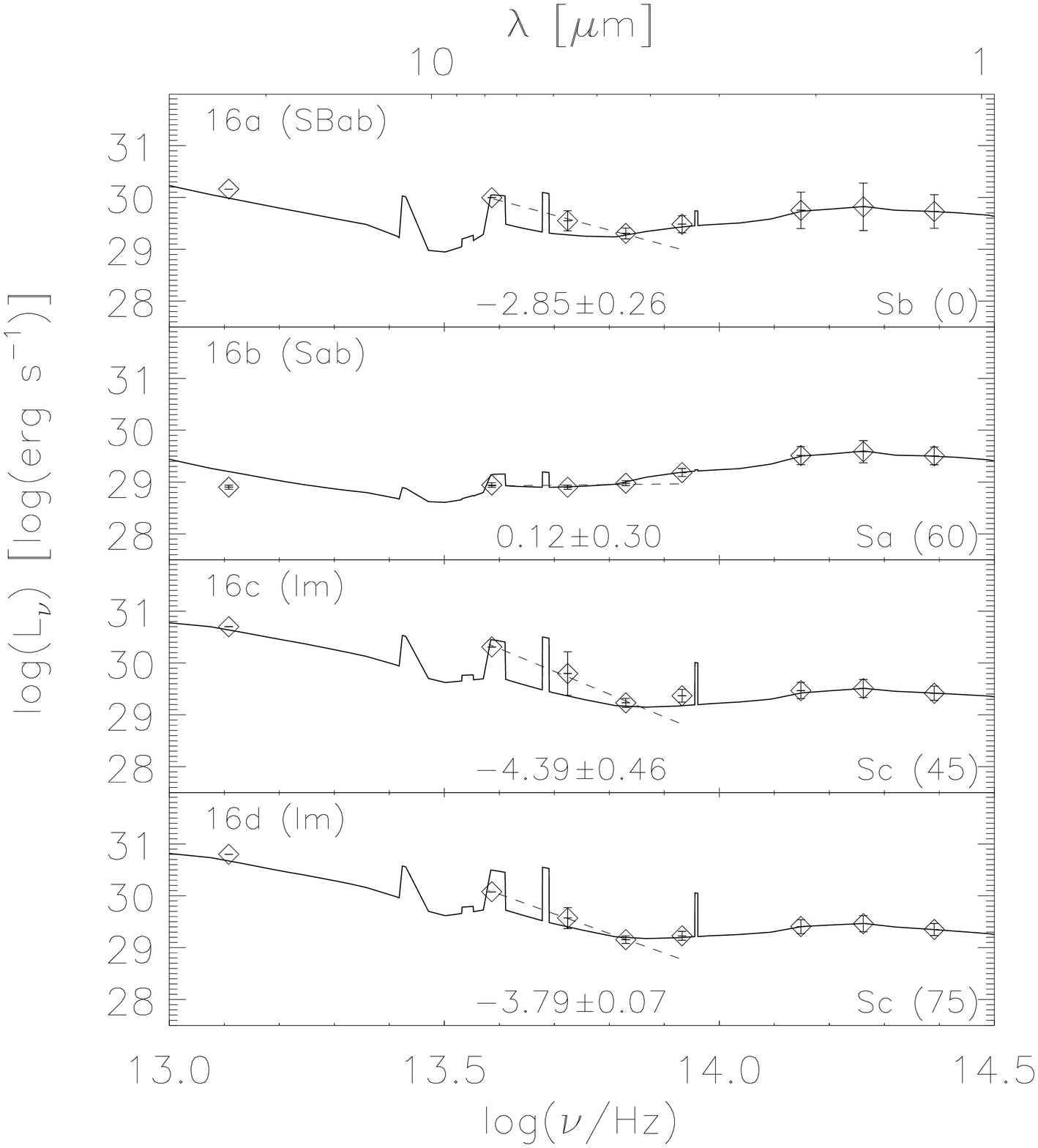}
		\includegraphics[width=0.329\textwidth,angle=0]{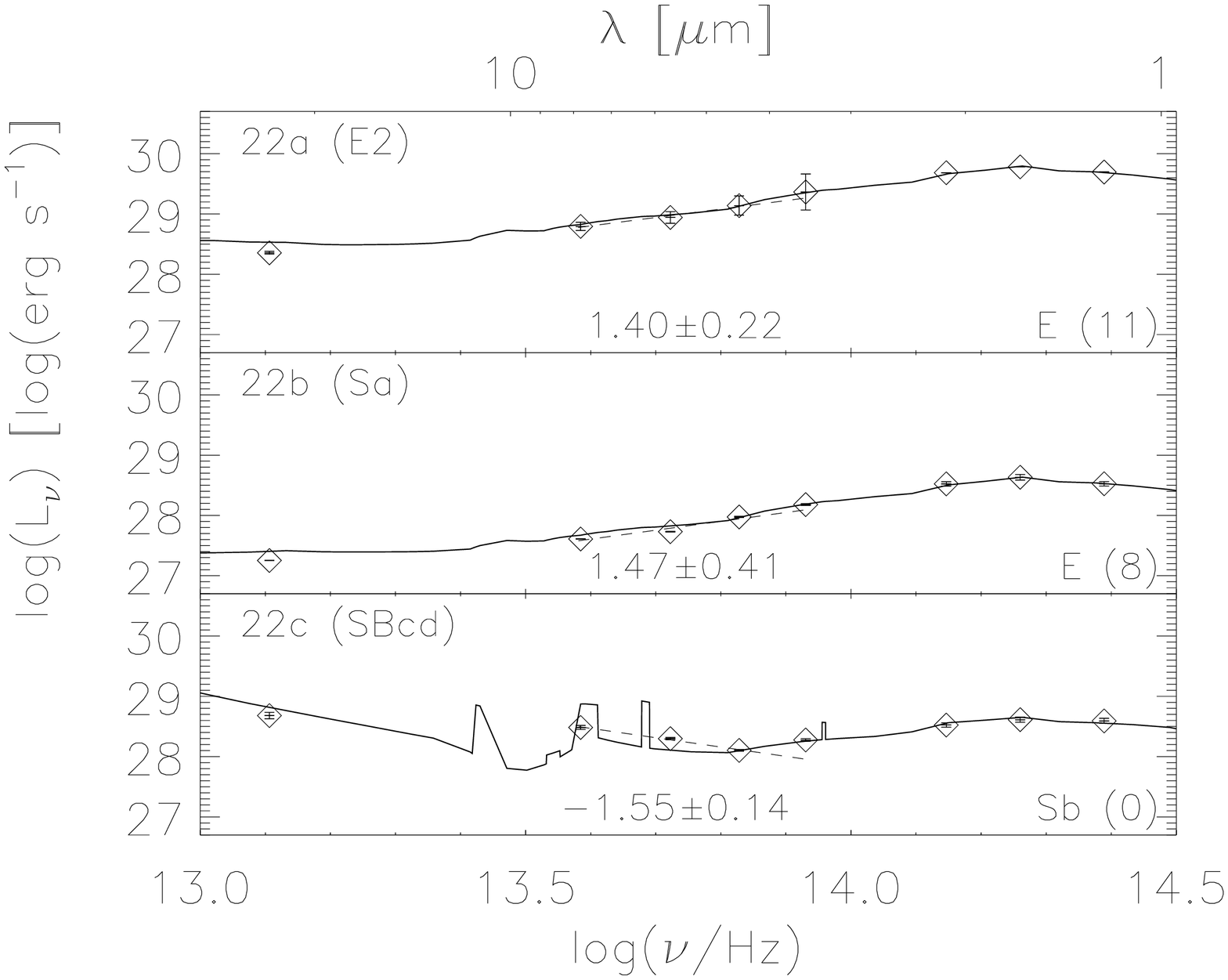}
		\includegraphics[width=0.329\textwidth,angle=0]{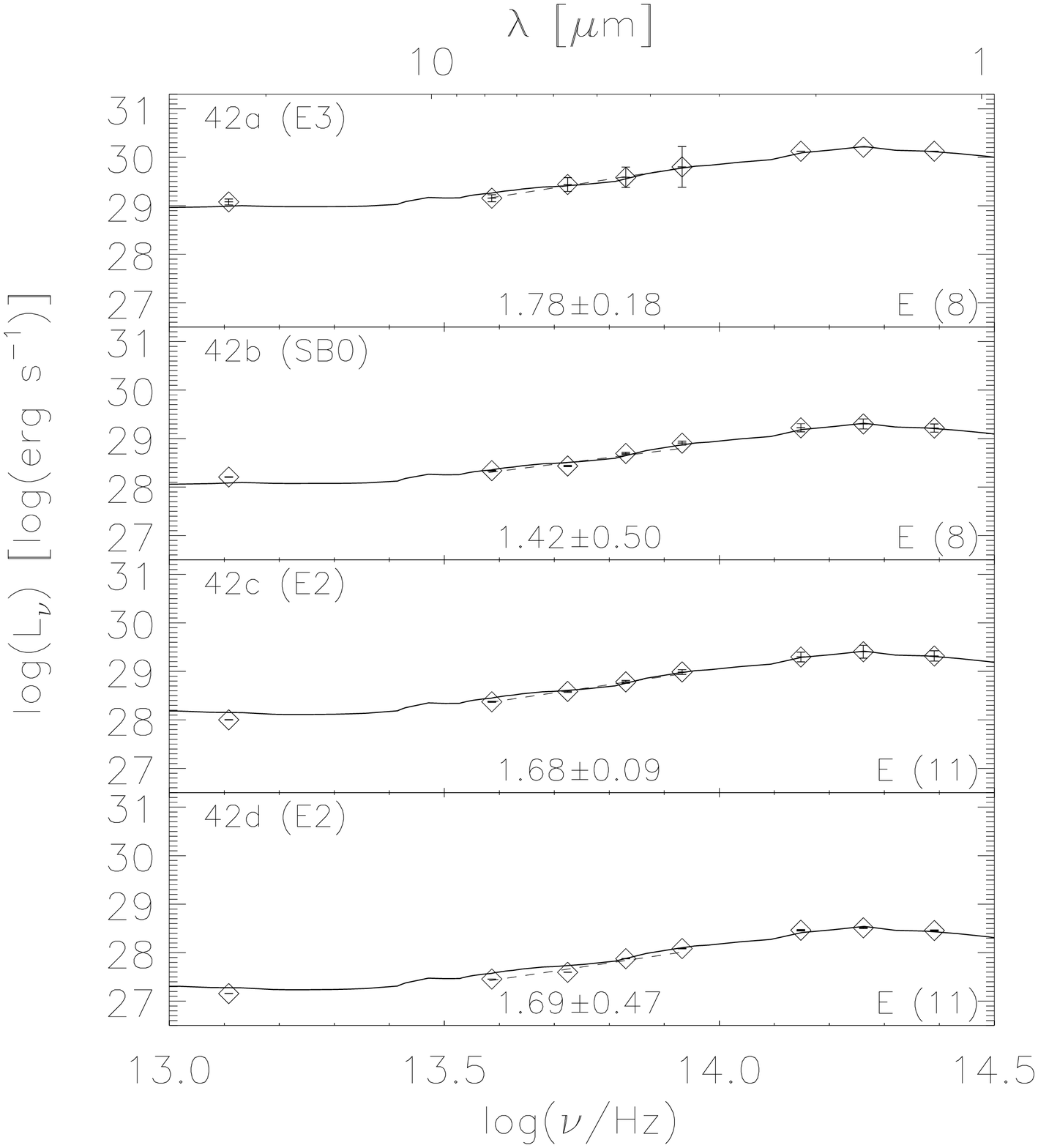}
		\caption{Infrared spectral energy distributions (SEDs) of 
			HCGs~16, 22, and 42 (left, center, right). The represented 
			galaxies follow the SEDs of their nominal morphological 
			types (marked in brackets next to the galaxy identifier),
			with the exception of HCG~22B, the galaxy highlighted in 
			Figure~\ref{fig:detail22}, right panel. The selected model 
			SEDs are indicated in the bottom right of each panel, 
			with the number in brackets indicating the logarithmic age 
			(yr) of early-type galaxy SEDs, or the inclination of 
			late-types. Numbers below the SED mark the \airac\ 
			diagnostic \citep{gallagher08} of mid-IR activity. This is 
			a power-law fit of the \spit-IRAC bands ($4.5-8.0~\mu$m), 
			therefore negative values denote activity, and positive ones 
			quiescence. 
		}\label{fig:seds}

	\end{center}
\end{sidewaysfigure*}

An overall image of normality is conveyed through the IR spectral energy distributions (SEDs) of the individual galaxies in all three groups, shown in Figure~\ref{fig:seds}. These follow the methodology of \citet{gallagher08} and combine 2MASS photometry in the JHK bands \citep{2mass}, the four \spit-IRAC bands (3.6, 4.5, 5.8, 8.0~$\mu$m), and $24~\mu$m from \spit-MIPS. Each plot lists the morphological type of a galaxy next to its identifier, followed by the morphology of the plotted GRASIL model \citep{silva98}. We also quote \airac~\citep{gallagher08}, a power-law fit to the $4.5-8.0$ part of the SED, which serves as a diagnostic of star formation activity: positive values denote quiescent galaxies, while star-forming systems register negative \airac. In previous works \citep[especially][]{isk10} we resorted to customizing the components of each GRASIL model in order to provide an adequate description of CG galaxies. Here, the use of `standard' GRASIL templates is sufficient. 

\subsection{Gas Content, Cool and Hot}\label{sec:gas}
The H\one~properties of the three groups are summarized in Table~\ref{tab:hi}, where we list \mhi, the evolutionary types according to the \citet{isk10} and \citet{vm01} classification schemes, and the \mhi\ deficiency according to \citet[][expressed as logarithmic mass in units of \Msun]{vm01}. The H\one\ morphologies are very different, from the common, extended envelope of HCG~16, to the single H\one-rich galaxy of HCG~22 (22C), to the low overall content of HCG~42. The \citet{isk10} scheme builds on the \citet{johnson07} ratios of gas-to-dynamical mass (I, II, III trace rich, intermediate and poor groups) and further divides groups according to the location of the gas: solely within galaxies (Type~A) or ones with an intra-group medium detectable in any wavelength (Type~B). In summary, HCGs 16, 22, and 42 represent Types IB, IIA, and IIIA, respectively.

\begin{table*}[tbh]
\begin{center}
\caption{HCGs~16, 22, 42: H\one\ Properties.}\label{tab:hi}
\begin{tabular}{lccccl} 
\hline
\hline
%
	\colhead{Identifier}                           & 
	\colhead{$\log(\mmhi/\Msun)$\tablenotemark{a}} & 
	\tmult{H\one~type\tablenotemark{b}}            & 
	\colhead{$\delta_\textup{\scriptsize H\one}$\tablenotemark{c}} & 
	\colhead{Structure}                            \\
						& 
						& 
\colhead{(K10)}			& 
\colhead{(VM01)}		& 
\colhead{$\log(\Msun)$}	& 
		\\
\hline
%
HCG~16  & 14.20  & IB                    & $2$ & $+0.41$ & envelope \\
HCG~22  &  9.13  & IIA                   & $2$ & $+0.55$ & single galaxy \\
HCG~42  &  9.40  & IIIA\tablenotemark{d} & $-$ & $-0.22$ & depleted \\
%
\hline
\end{tabular}
\tablenotetext{a}{H\one\ masses from \citet[][HCG~16]{borthakur10}, 
					\citet[][HCG~22]{price00}, 
					\citet[][HCG~42]{huchtmeier94}.}
\tablenotetext{b}{Classifications by \citet{isk10}, \citet{vm01}.}
\tablenotetext{c}{Deficiency in H\one\ mass, as compared to field 
					galaxies of matched morphological types. After
					\citet{borthakur10} and \citet{vm01}.}
\tablenotetext{d}{The type of HCG~42 is uncertain, as it is not clear 
					whether the diffuse X-ray can be attributed to 
					the IGM or galaxy 42A alone. }
\end{center}
\end{table*}

The properties of the hot gas in the three groups are not quite as diverse as those of the H\one. The detailed analysis of \citet{desjardins13} detects no significant hot IGM component in the three groups, albeit the \chan\ data of HCG~16 are perhaps too shallow to make that assessment. Diffuse X-ray emission is, however, detected around certain galaxies. The region around 42A resembles an extended envelope of hot gas, similar in appearance to those found around massive galaxies in clusters \citep[\eg][]{ponman99}. In addition, data from the Survey for Ionization in Neutral-Gas Galaxies \citep[SINGG][]{meurer06} show that the hot gas plumes observed in X-rays are coincident with H$\alpha$ emission, as noted by \citet{werk10}. Finally, \citet{jeltema08} presented some tentative evidence for an X-ray bridge connecting 16A and B in Chandra imaging.

\section{Stellar Populations}\label{sec:stellarpops}

\subsection{Young and Intermediate-Age Star Cluster Populations}\label{sec:clusters}
In previous installments of this series we have used star cluster populations to add to the characterization of CGs. They are particularly helpful in accounting for star formation in cases where our broad-band metrics are contaminated by AGN (a consideration in this case; see Table~\ref{tab1}). Here we take advantage of the opportunity to contrast the populations of three groups representing a three-stage sequence along the \citet{isk10} evolutionary diagram: types IB, IIA, IIIA. 

\begin{figure*}[tbhp]
	\begin{center}

		\includegraphics[width=0.49\textwidth,angle=0]{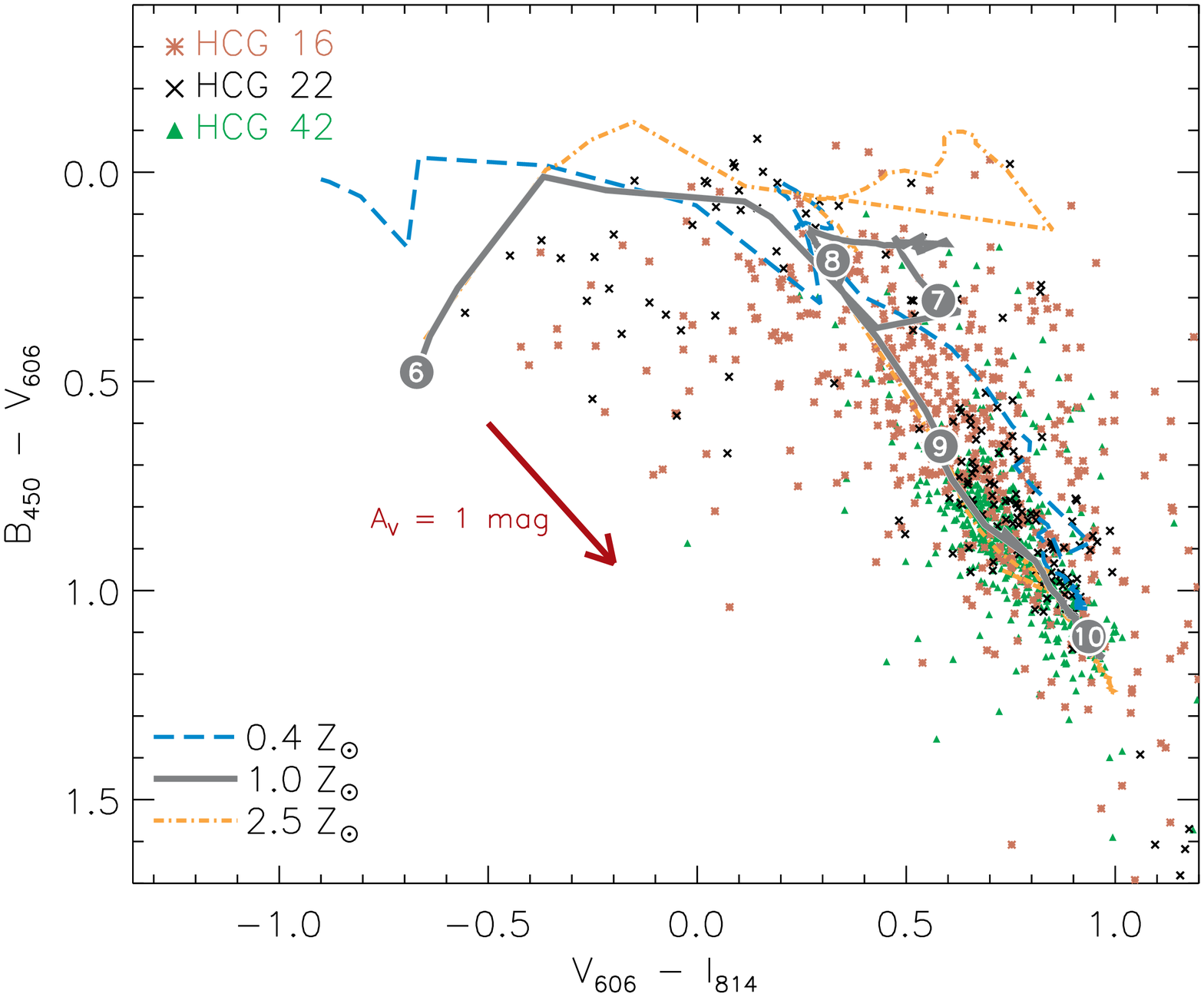}
		\includegraphics[width=0.49\textwidth,angle=0]{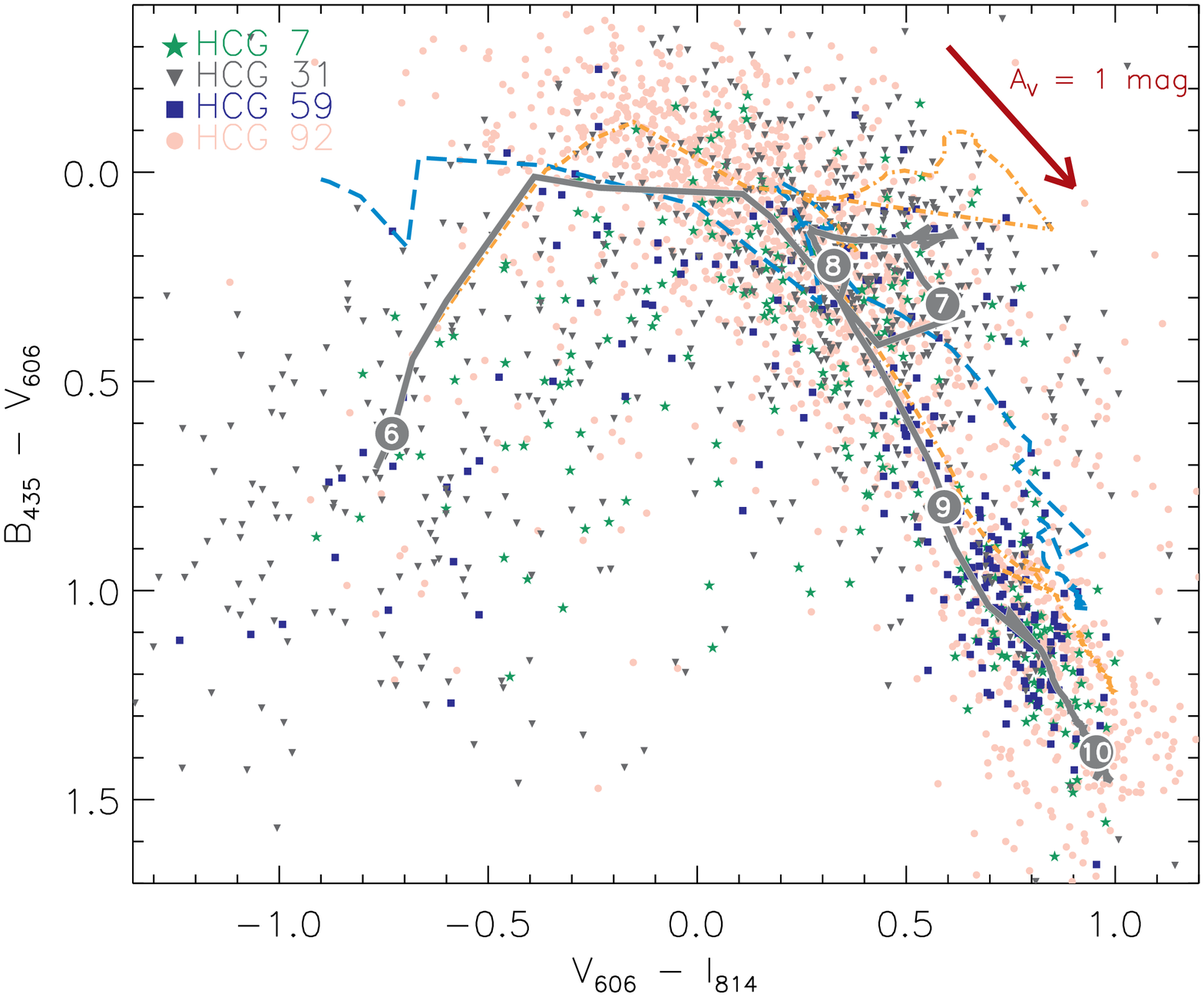}
		\caption{Star cluster candidates in HCGs~16, 22, 42 
			(\textbf{left}, WFPC2 filters), and the combined 
			populations of four other groups, shown on the 
			\textbf{right} as a benchmark (ACS filters). The 
			dashed, solid, and dashed-dotted lines trace the 
			evolution of \ygg\ SSP models of $[0.4, 1.0, 2.5]~\Zsun$
			\citep{zackrisson11}. The model tracks include 
			nebular emission transmitted in the F435W and 
			F606W filters at ages $\lesssim7~$Myr. The extinction 
			vector has a length of 1 magnitude ($A_V$). Sources 
			aligned with vectors originating on the nebular 
			segment have ages less than $\approx7~$Myr. The 
			relative dearth of sources in the nebular part of 
			color space, as compared to all HCGs of the right 
			panel, diagnoses relatively weak current star cluster 
			formation activity throughout. 
			The three groups display different star formation 
			histories. HCG~16 features many clusters of super-solar 
			metallicity, while it does not appear to host a 
			distinct population of old globular clusters. 
			This is to be expected, given the low galaxy masses 
			and mainly late-type morphologies. Both HCG~22 
			and 42 show remarkable populations, dominated by 
			clusters at intermediate and old ages ($>1~$Gyr), 
			with only very few young clusters in HCG~22C. This 
			is probably related to a group-wide era of merger-driven 
			star formation in the past, consistent with the 
			morphological types of the galaxies in these two 
			groups, and the various signs of disturbance. 
			From this we diagnose that HCGs~42, and two-thirds
			of HCG~22 entered a different mode of evolution 
			(with little conversion of gas to stars) over the 
			past $\sim1~$Gyr.
		}\label{fig:cc}

	\end{center}
\end{figure*}

This sequence is largely reflected in the star cluster populations, displayed in Figure~\ref{fig:cc}. Here we make use of \ygg\ SSP models, as they incorporate nebular emission and continuum transmitted in the \bb\ and \vb\ filters\footnote{In Figure~\ref{fig:agemass} the nebular features were not desirable, hence we used the SSPs of \citet{marigo08}.}. Star clusters can be roughly age-dated through this diagram by comparing their position in color space to the SSP track, which evolves from 6~Myr to $\sim10$~Gyr. Specifically, HCG~16 shows evidence of star formation extending to a few Gyr into the past, with much activity over the past Gyr, consistent with star formation in the three late-type galaxies. It does not, however, show a pronounced globular cluster (GC) clump at ages $\sim10~$Gyr, as one should expect from the lack of elliptical galaxies and the low overall masses of the galaxies. 

In HCG~22 we detect few young clusters, with most sources found to have ages older than 1~Gyr. This denotes little activity over the past Gyr (centered around the spiral galaxy 22C, as is the H\one), and pronounced star formation before that mark. It is therefore evocative of a gas-depleted system, in a fashion somewhat contrary to its \citet{johnson07}/\citet{isk10} type. While it features two early-type galaxies among three members, it only shows a weak GC clump. While the specific frequency for GC-rich HCG~22A is consistent with its morphological type (as discussed in Section~\ref{sec:literature}), the GC clump is rather weak. This is rather interesting, as galaxies are not normally found to have middle-age-heavy cluster populations (\ie\ between 1 and a few Gyr old), hence HCG~22 seems to have undergone a period of intense star formation in the past ($>1~$Gyr). The hints of post-merger morphologies in galaxies 22A, 22B might offer a clue as to the origin of this notable star cluster population. 

At the evolved end of the gas-richness sequence, HCG~42 shows little evidence of star formation in the past $\sim$Gyr, with a pronounced GC clump, a distribution consistent with its IIIA evolutionary type. In this case, the areas of color space that are void are perhaps more interesting than interpreting those that are full: the absence of star clusters at ages younger than $1~$Gyr indicates that star formation was essentially switched off at that point in time. This is when HCG~42 seems to have entered a different phase of evolution, one featuring little conversion of gas into stars. 

In contrast with our previous studies of HCGs~7, 31, and 59~\citep{gallagher10,isk10,isk12a}, we find no large-scale star cluster complexes in these three groups (except perhaps in the spiral arms of HCG~16A). This should be expected in quiescent HCGs~22 and 42, and possibly explained by the apparently high dust content in star-forming galaxies 16C and 16D.

\subsection{The Globular Cluster Population of HCG~42A}\label{sec:gc}
The old globular clusters (GCs) are more difficult to study, due to the bright limiting magnitude and restricted field coverage of the WFPC2 imaging. While HCG~22A (NGC 1199) does harbor a significant GC population \citep{barkhouse01}, the number of GCs present in our WFPC2 data is not large enough to allow a more detailed study. 

We do, however, detect a very large population of GCs in our ACS images of HCG~42A, a massive elliptical galaxy. We alter the source selection cuts of Section~\ref{sec:select} to allow for the inclusion of more clusters. This is acceptable here, as GCs have very tightly confined colors. We therefore adopt a limiting magnitude of $V_{606}=25.2$, or $M_V=8.7~$mag at the adopted distance of $59~$Mpc. We thus sample the top 15$\pm$5\% of the GC luminosity function, assuming a luminosity function turnover at $M_V=-7.2\pm 0.2~$mag \citep[\eg][]{jordan07}. 

\begin{figure*}[bt]
	\begin{center}

		\includegraphics[height=180pt,angle=0]{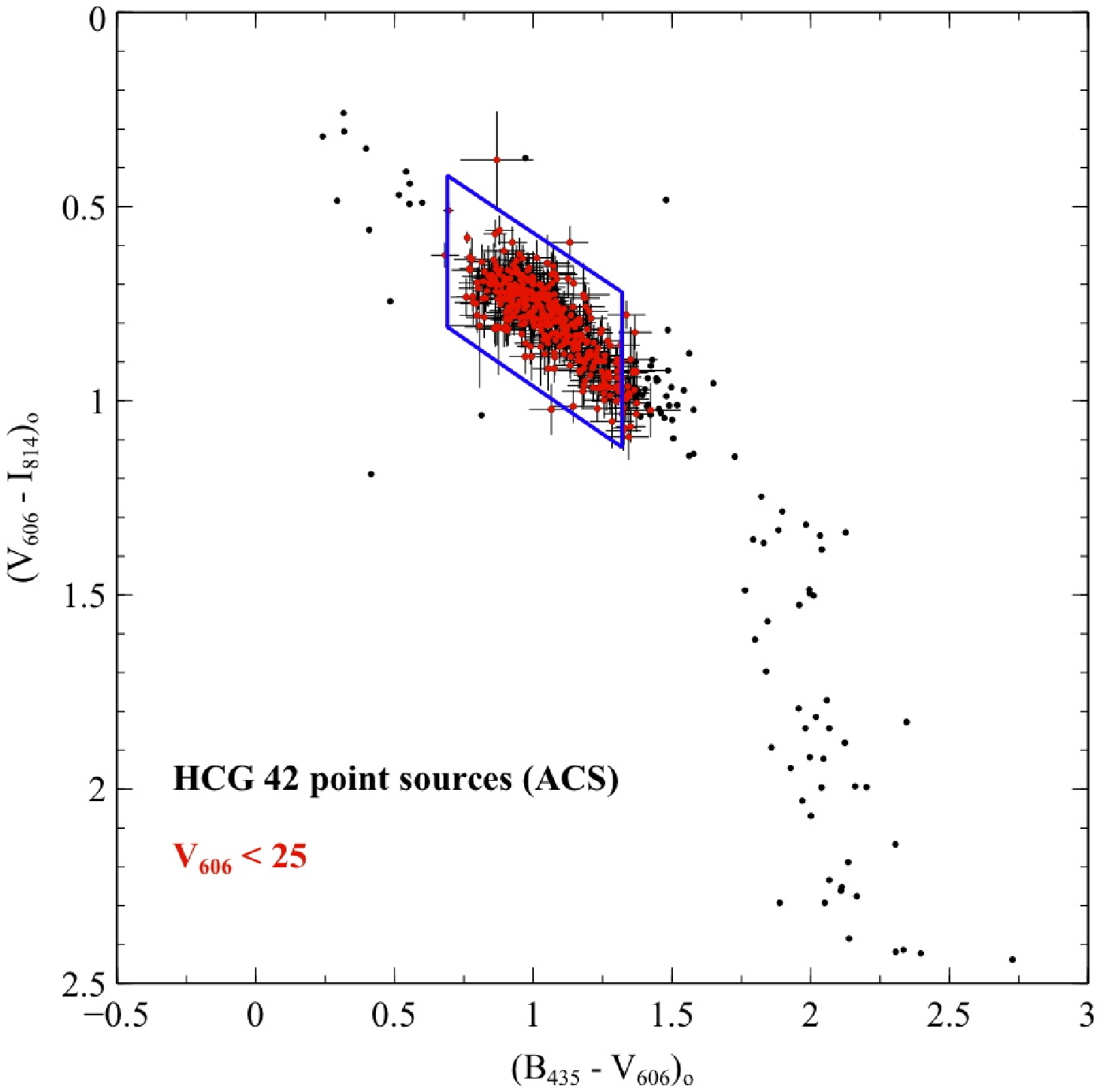}
		\includegraphics[height=180pt,angle=0]{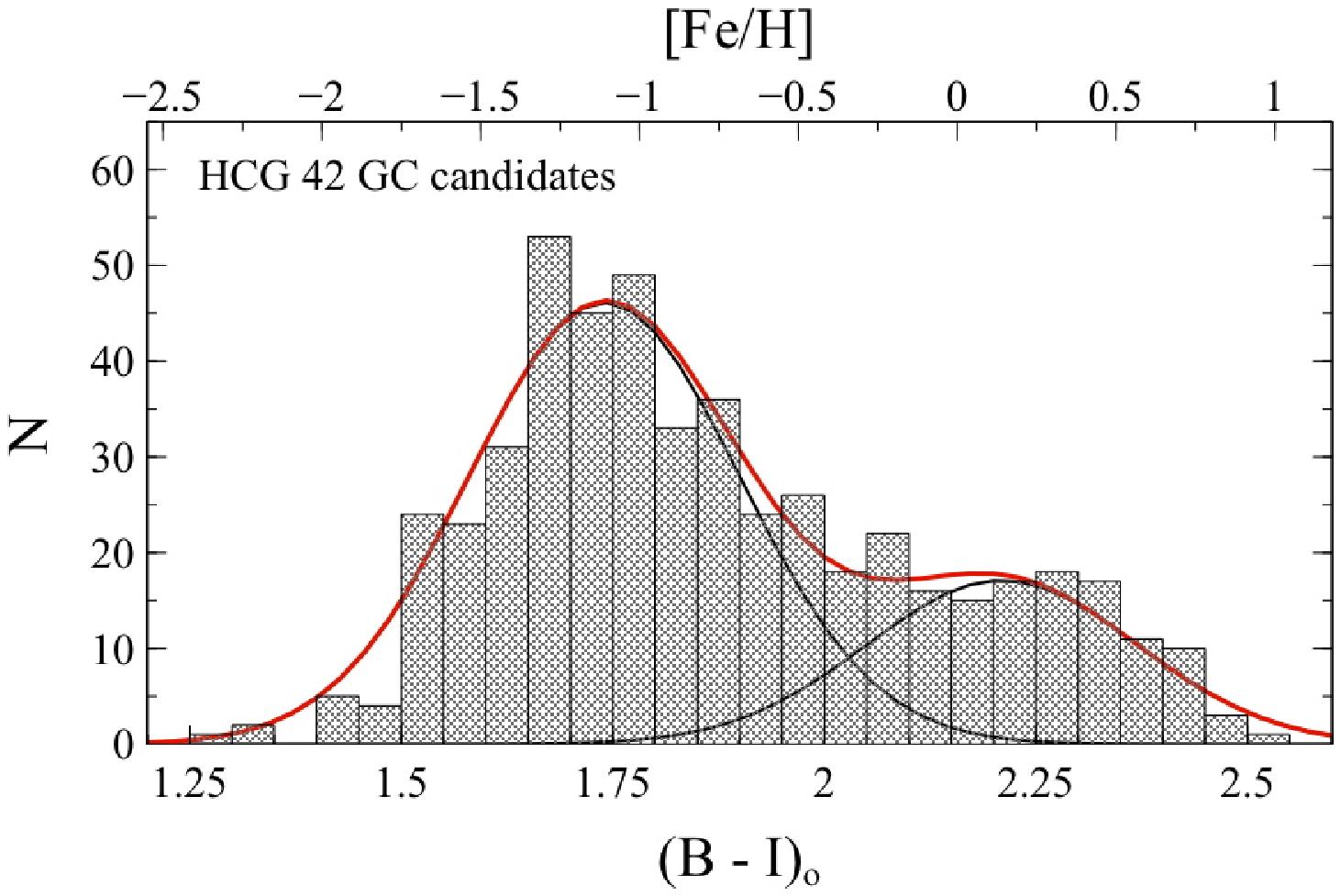}
		\caption{Globular clusters in HCG~42. 
		Globular cluster selection (left) and color/metallicity
		distribution for HCG~42. The brightness cut at 
		$m_{606} < 25~$mag corresponds to $M_{606} < -8.8~$mag, 
		which is brighter than the peak of the GC luminosity 
		function \citep[\eg][]{jordan07}, which complicates our
		interpretation. We find a bimodal distribution, common 
		among early-type galaxies in clusters but not necessarily 
		those in CGs \citep[\eg][]{isk10,isk12a}.
		}\label{fig:gc}

	\end{center}
\end{figure*}

We detect a total of 489 objects with GC-like colors, shown in the color-color diagram of Figure~\ref{fig:gc} (left). The $B_{435}-$\vb\ colors from the ACS photometry were converted to Johnson-Cousins $B-I$ colors using the synthetic transformations from \citet{sirianni05}. These colors were in turn converted to the metallicity [Fe/H] using the conversion relation from \citet{harris06} and gave rise to the metallicity distribution of Figure~\ref{fig:gc} (right). The histogram shows a clear bimodal distribution. An analysis according to the  KMM metric of \citet{ashman94} reveals the peak of the `blue' globular cluster distribution at $(B-I)_0=1.74~$mag, or [Fe/H]$=-1.2$, and that of the `red' globular clusters at $(B-I)_0=2.21~$mag, or [Fe/H]$=+0.1$. Related uncertainties are expected to arise from the photometric calibration and the color-to-[Fe/H] conversion, at the $0.3~$dex level.

We are also able to extrapolate the total number of GCs in HCG~42A by correcting the number of detections within the ACS frame to the nominal area of the entire galaxy. We derived the radial profile of the GC system in a series of elliptical annuli with a fixed $\epsilon =0.25$. Since our single ACS frame does not cover a background region, we used imaging of HCG~7 \citep[from][at a similar distance of 65~Mpc]{isk10} to estimate background contamination in each annulus, to the edge of the ACS frame (a major axis distance of 3.7\arcmin). Based on Poisson errors in the number counts, and considering background contamination at the level of $31\pm 11$ GCs, we predict a total number of clusters above the cutoff magnitude of $N_\textup{\scriptsize GC}^\textup{\scriptsize bright}=680\pm63$. Adopting a completeness level of $0.9\pm 0.1$ and extrapolating to the entire GC luminosity function, we derive a total GC population of HCG 42A of $N_\textup{\scriptsize GC}=5030\pm2140$. Combined with the absolute magnitude of $M_V=-23.2~$mag\footnote{Based on the integrated $R$ magnitude in Table 1 and assuming $V-R=0.5~$mag.} for the host galaxy yields a specific frequency $S_N =2.6\pm 1.1$ (the number of clusters per unit luminosity). All the above properties of HCG 42A, its morphology, specific frequency, GC bimodality, and location of color/metallicity peaks, are very similar to those of luminous galaxies in the cluster environment \citep[\eg][]{harris06, brodiestrader06, peng06}. These attributes could be related with HCG~42 being embedded in a larger structure, as will be discussed in the following section. 

\subsection{Dwarf Galaxies}\label{sec:dwarfs}
We now turn our attention to the dwarf galaxy populations of the three HCGs. We combine information from the literature and new Hydra spectroscopy to study a total of 59 dwarf galaxies. More specifically, we incorporate the datasets of \citet[][hereafter dC97]{decarvalho97}, \citet[][ZM00]{zm00}, and \citet{carrasco06}, albeit for only a limited projected area about the group center. It is worth noting that the \citeauthor{carrasco06} catalogue builds on the previous ones and extends coverage to faint targets (down to $R\approx21$). We also make use of previously unpublished velocity tables from the spectroscopic campaign that provided the basis for the \citet{zm98,zm00} papers on poor groups (referred to as ZM98 throughout this section). Given the inhomogeneity of these collated catalogues, it is challenging to fully assess the completeness both in space and brightness. The samples are, however, complete to $R\approx18~$mag, as they draw from the ZM00 catalogues (see their Figure~3), conditional on no galaxies having been filtered out by the ZM00 selection. Given the mean co-moving radial distances to the three groups \citep[based on redshifts from][]{hickson92}, this detection limit corresponds to $M_R=[-15.5, -14.7, -15.9]~$mag. The \citeauthor{carrasco06} sample extends the completeness of the HCG~42 dwarf galaxy catalogue to $V=20~$mag at the 80\% level ($M_V=-13.9~$mag) over a limited central area. 

\begin{figure*}[tbhp]
	\begin{center}

		\includegraphics[width=0.32\textwidth,angle=0]{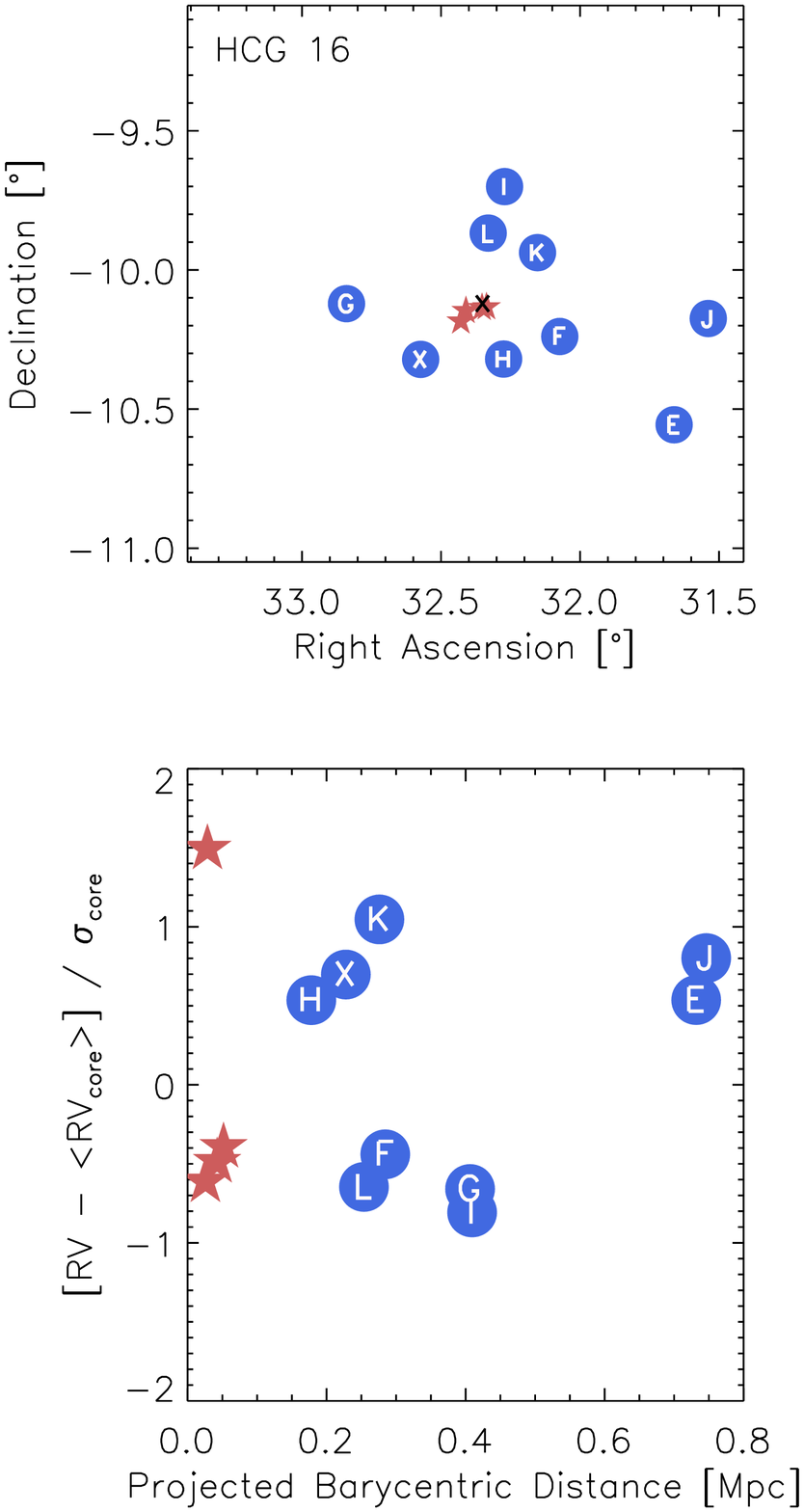}
		\includegraphics[width=0.32\textwidth,angle=0]{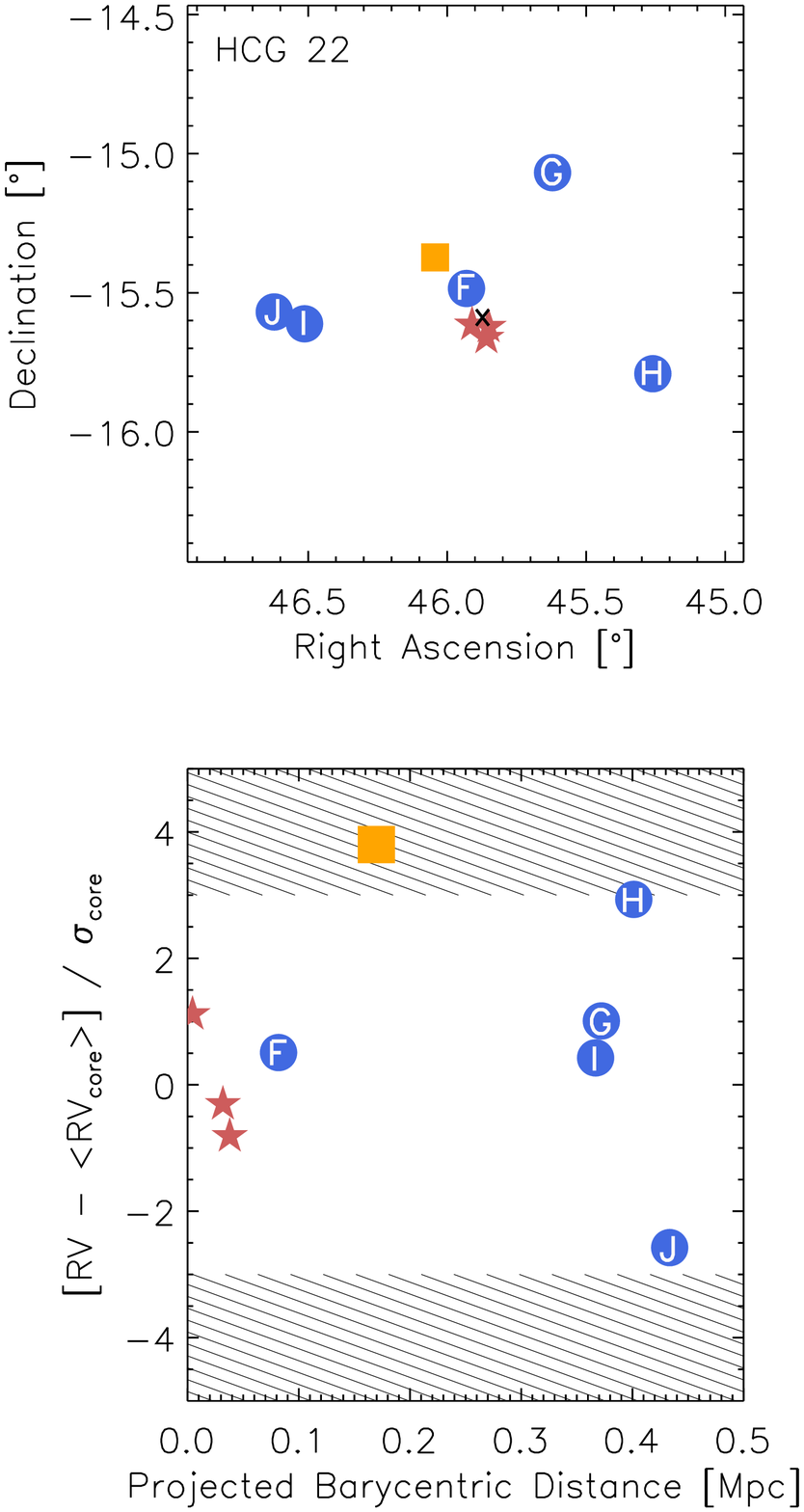}
		\includegraphics[width=0.32\textwidth,angle=0]{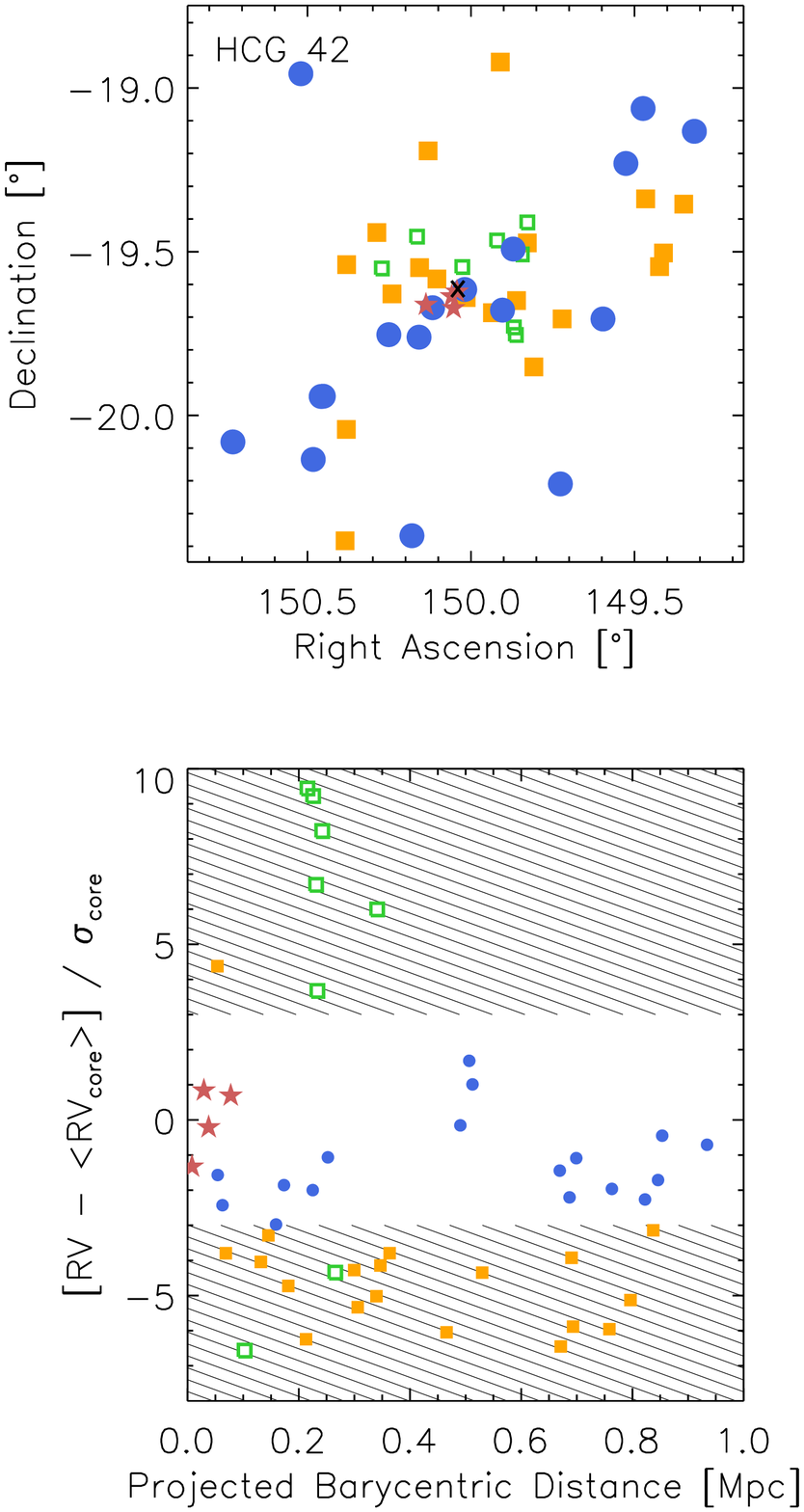}
		\caption{Diagnostic maps (\textbf{top row}) and phase-space 
			diagrams (\textbf{bottom row}) for candidate member galaxies 
			in HCGs~16, 22, and 42 (left, center, right). Main 
			members are marked with stars and their barycenter as a 
			cross. Lettered circles and yellow boxes mark galaxies 
			that qualify as members and associates (possible members, 
			see Section~\ref{sec:dwarfs} for details). 
			The phase-space diagrams plot the projected distance 
			from the group barycenter against the deviation in 
			velocity,  in terms of $\sigma$ away from the main 
			member mean. The hashed regions mark the $3\,\sigma$ 
			boundary that qualifies a galaxy as a group member. 
			HCG~16 appears to be a populous group, with seven 
			lesser members in addition to the four main giants. 
			Galaxy~X lies at the end of the greater H\one\ cloud 
			that covers the main members \citep{vm01}. 
			HCG~22 shows five member dwarfs and one associate 
			according to our employed diagnostic. The $3\,\sigma$ 
			cut is, however, based on a small number of 
			main galaxies, hence the associate might in reality 
			be a member. 
			HCG~42 is a relatively rich group, numbering 17 
			members and 26 associates. The open green boxes mark
			galaxies studied by \citet{carrasco06} and not included
			in any previous catalog. They cover a fainter part
			of the luminosity function than \citet{zm00} and 
			\citet{decarvalho97}, albeit in a limited field of side 
			50\arcmin. The velocity structure appears to place 
			the giants at the end of a large structure, perhaps a 
			filament. 
		}\label{fig:pspace}

	\end{center}
\end{figure*}

Figure~\ref{fig:pspace} plots all galaxies within $5\,\sigma$ of the velocity dispersion for each group, derived from giant galaxies only. This was adapted to $10\,\sigma$ for HCG~42 since it is situated in a galaxy-rich region, although we note that the majority of these galaxies lie within the $5\,\sigma$ velocity cut. Given the large error associated with deriving a statistical dispersion from so few datapoints, anything in this $5\,\sigma$ velocity range is considered an `associate', while we enforce a $3\,\sigma$ criterion for membership. Associates and members are plotted as orange boxes and blue circles respectively, while the \citet{carrasco06} galaxies are plotted as open green boxes, to mark their different completeness level and spatial coverage. The following sections visit each group individually. 

\subsubsection{HCG~16}\label{sec:dwarfs16}
HCG~16 is covered by SDSS, so we performed a spectroscopic search in a degree-wide area centered on the group barycenter (the mass-weighted mean RA and Dec). We allow for a spread of $\pm5~\sigma$ about the mean redshift, representing the above-mentioned selection of group associates. The search yields four galaxies beside the four group giants, none of which were previously known to be associated with HCG~16. Furthermore, the unpublished ZM98 tables list three objects at accordant redshifts, and two galaxies from the dC97 catalogue complete a set of nine dwarf members down to $M_R\approx18~$mag. They are distributed about the giant galaxies in a roughly symmetric fashion, inside a projected area of side 0.8~Mpc. Interestingly, all these dwarfs apart from 16X lie outside the shared H\one\ envelope of the HCG~16 giants. This might suggest that the gas originated in the individual giants and was released through interactions. The galaxies are listed in order of decreasing brightness on Table~\ref{tab:dwarfs_1622}, derived from ZM98 photometry in the B-band (except the dC97 objects, where dC97 B-band brightness is listed). We use the \citet{hickson82} naming convention to assign letters to these galaxies from E to L (assuming completeness to the faintest member). Incorporating the velocities of these nine objects increases the dynamical mass of the group by a factor of $\approx4$. 

\begin{deluxetable}{clccccrc}
\tabletypesize{\small}
\tablewidth{0pt}
\tablecolumns{8}
\tablecaption{Galaxies at Accordant Redshifts with HCGs~16 and 22.\label{tab:dwarfs_1622}}
\tablehead{
	&\colhead{ID}								& 
	\tmult{Coordinates}						& 
	\colhead{Brightness}	& 
	\colhead{$v_R$}							& 
	\colhead{$d_\textup{\scriptsize BC}$}	& 
	\colhead{Morphology}					\\ 
	%
 	&								& 
	\colhead{(h~m~s)}						& 
	\colhead{($^\circ~^\prime~^{\prime\prime}$)} & 
	\colhead{(mag)}							& 
	\colhead{(\kms)}						& 
	\colhead{(kpc)}							& 
}
\startdata
%
\sidehead{\textbf{HCG~16}}
X & \n848   								& 02~10~17.6 & -10~19~17.0 & 13.60 & 3989 &  0.23 & SBab pec \\
E & APMUKS(BJ) B020413.18-104749.8			& 02~06~38.8 & -10~33~23.2 & 15.56 & 3830 &  0.41 & $-$\\
F & KUG~0205-104							& 02~08~17.5 & -10~14~21.0 & 15.78 & 3869 &  0.28 & $-$ 			\\
G & KUG~0208-103							& 02~11~21.6 & -10~07~16.3 & 15.89 & 3846 &  0.41 & $-$ 			\\
H & KUG~0206-105							& 02~09~06.0 & -10~19~13.0 & 16.35 & 3971 &  0.18 & Sc 			\\
I & KUG~0206-099A 							& 02~09~05.2 & -09~42~01.4 & 16.77 & 3830 &  0.41 & $-$ 			\\
J & hcg~16--10								& 02~08~36.8 & -09~56~16.6 & 17.52 & 4000 &  0.18 & $-$\\ 
K & SDSS J020836.69-095615.7				& 02~08~36.7 & -09~56~15.8 & 17.63 & 4025 &  0.28 & $-$			\\
L & SDSS J020919.31-095202.0 				& 02~09~19.3 & -09~52~03.3 & 19.79 & 3847 &  0.25 & S 				\\
\sidehead{\textbf{HCG~22}}
F & \n1209 									& 03~06~03.1 & -15~36~41.8 & 12.84 & 2690 &  0.38 & E6? 			\\
G & \n1188 									& 03~03~43.4 & -15~29~04.5 & 14.12 & 2698 &  0.09 & SAB0 \\
H & \n1231	 								& 03~06~29.3 & -15~34~08.4 & 14.32 & 2424 &  0.45 & Sc				\\
$-$& APMUKS(BJ) B030149.90-153401.7			& 03~04~10.4 & -15~22~22.1 & 16.55 & 2991 &  0.19 & $-$			\\
I & APMUKS(BJ) B030008.22-151549.8 			& 03~02~29.0 & -15~04~05.8 & 17.19 & 2743 &  0.37 & $-$			\\
J & APMUKS(BJ) B025842.23-155916.7			& 03~01~02.4 & -15~47~27.7 & 17.29 & 2913 &  0.38 & $-$			\\
\enddata
\tablecomments{Coordinates and radial velocities are drawn from previously unpublished catalogues 
					related to \citet{zm98}, apart from: hcg~16--10 
					\citep[][coordinates converted from B1950 to J2000]{decarvalho97}; 16G, I, K, and L \citep{sdss}.
					We quote SDSS $g$-band magnitudes for HCG~16 member galaxies, with the exception of
					hcg~16--10, which is a B magnitude from \citet{decarvalho97}. HCG~22 values 
					are POSS photographic magnitudes \citep{poss}.
					$d_\textup{\scriptsize BC}$ is the distance of a dwarf from the group barycenter.
					The morphologies of some galaxies not in the New General Catalogue were drawn 
					from the catalogues related to \citet{zm98}. The majority were not classified, 
					owing to the limitation of available imaging. In the case of NGC galaxies we 
					refer to \citet{rc3}. 
					}
\end{deluxetable}

\subsubsection{HCG~22}\label{sec:dwarfs22}
In HCG~22 the situation is similar, with five members and one associate, which we draw from the ZM98 tables. We employ the \citet{hickson82} lettering scheme for members only. It should be noted here that the dispersion is derived from three velocities, therefore the membership of this galaxy is difficult to assess. The inclusion of just members, or all dwarfs (\ie\ members and the one associate) increase the \mdyn\ by a factor of 30 or 50 respectively.

\subsubsection{HCG~42}\label{sec:dwarfs42}
The third group, HCG~42, presents an altogether different image, with a large number of member and associate galaxies spread out widely in three dimensions. We consider 34 dwarfs from ZM00, eight from \citet{carrasco06}, two from dC97, and one from our Hydra redshift survey. These are split into 35 associates and 17 members in a relatively small projected area of $1~$Mpc$^2$. Interestingly, the phase-space distribution of Figure~\ref{fig:pspace} (right) places the four giant galaxies at the top end of the bright galaxy $\sigma$ distribution -- recall that the \citet{carrasco06} systems, marked as green boxes, are fainter than the rest. With virtually all other associates at lower redshifts, this could imply that HCG~42 is part of a larger structure that may extend to the foreground. The substructure echoes the theoretical prediction of \citet{mcconnachie09cg} and observational assessment of \citet{mendel11} who find that $\sim50\%$ of all CGs are embedded in larger structures. 

\begin{deluxetable}{rccccrcc}
\tabletypesize{\footnotesize}
\tablewidth{0pt}
\tablecolumns{8}
\tablecaption{Galaxies at Accordant Redshifts with HCG~42.\label{tab:dwarfs_42}\vspace{-10pt}}
\tablehead{
	\colhead{ID}							& 
	\tmult{Coordinates}						& 
	\colhead{Brightness} 					& 
	\colhead{$v_R$}							& 
	\colhead{$d_\textup{\scriptsize BC}$}	& 
	\colhead{Morphology}					\\
													& 
	\colhead{(h~m~s)}								& 
	\colhead{($^\circ~^\prime~^{\prime\prime}$)}	& 
	\colhead{(mag)}									& 
	\colhead{(\kms)}								& 
	\colhead{(kpc)}									& 
}
\startdata
$[$ZM00$]$~0003 & 10~00~43.3 & -20~22~03.6 & 13.60 &  3977 &  0.70 & $-$ \\
$[$ZM00$]$~0005 & 09~59~29.0 & -19~29~30.7 & 14.07 &  3853 &  0.69 & $-$ \\
$[$ZM00$]$~0015 & 10~01~48.3 & -19~56~29.7 & 14.64 &  4081 &  0.49 & $-$ \\
$[$ZM00$]$~0017 & 09~58~06.3 & -19~13~49.6 & 14.75 &  4049 &  0.85 & $-$ \\
$[$ZM00$]$~0022 & 10~00~28.2 & -19~40~15.9 & 15.04 &  3879 &  0.76 & SB0 \\
$[$ZM00$]$~0029 & 10~00~38.1 & -19~45~40.0 & 15.30 &  3766 &  0.16 &dE0N \\
$[$ZM00$]$~0033 & 10~02~04.8 & -18~57~22.6 & 15.57 &  4212 &  0.51 & $-$ \\
$[$ZM00$]$~0034 & 10~01~00.3 & -19~45~12.5 & 15.67 &  3980 &  0.25 & S0  \\
$[$ZM00$]$~0041 & 09~57~16.1 & -19~07~56.1 & 15.83 &  3891 &  0.17 & $-$ \\
$[$ZM00$]$~0046 & 09~59~36.9 & -19~40~42.7 & 16.02 &  4020 &  0.93 & dE  \\ 
$[$ZM00$]$~0057 & 10~02~54.7 & -20~04~52.1 & 16.22 &  3938 &  0.67 & $-$ \\
$[$ZM00$]$~0058 & 09~58~54.4 & -20~12~34.0 & 16.28 &  3846 &  0.82 & $-$ \\
$[$ZM00$]$~0065 & 10~01~55.8 & -20~08~05.0 & 16.43 &  3908 &  0.85 & $-$ \\
$[$ZM00$]$~0106 & 10~00~04.6 & -19~36~55.4 & 16.88 &  3876 &  0.23 & dE  \\ 
$[$ZM00$]$~0143 & 09~58~23.1 & -19~42~19.9 & 17.23 &  4287 &  0.51 & S\ldots/Irr  \\
$[$ZM00$]$~0154 & 10~01~49.9 & -19~56~32.0 & 17.30 &  3924 &  0.05 & $-$ \\
$[$ZM00$]$~0166 & 09~57~53.7 & -19~03~45.0 & 17.36 &  3828 &  0.06 & S   \\
\tableline
$[$ZM00$]$~0002 & 09~57~23.9 & -19~21~16.9 & 12.99 &  3675 &  0.07 & dE0N\\
$[$ZM00$]$~0006 & 10~01~31.2 & -19~32~22.3 & 14.13 &  4587 &  0.05 & $-$ \\
$[$ZM00$]$~0009 & 10~00~31.5 & -19~11~30.7 & 14.42 &  3675 &  0.36 & $-$ \\
$[$ZM00$]$~0013 & 10~01~09.1 & -19~26~29.1 & 14.51 &  3442 &  0.69 & SBd \\
$[$ZM00$]$~0014 & 09~59~13.9 & -19~51~07.8 & 14.59 &  3526 &  0.80 & S0  \\
$[$ZM00$]$~0016 & 09~57~38.8 & -19~30~14.1 & 14.67 &  3424 &  0.47 & $-$ \\
$[$ZM00$]$~0019 & 10~01~32.4 & -20~23~00.0 & 14.82 &  3504 &  0.31 & SBab\\
$[$ZM00$]$~0021 & 09~58~53.1 & -19~42~19.1 & 14.85 &  3621 &  0.30 & $-$ \\
$[$ZM00$]$~0023 & 10~01~31.4 & -20~02~34.9 & 15.04 &  3732 &  0.15 & $-$ \\
$[$ZM00$]$~0026 & 09~57~51.8 & -19~20~18.9 & 15.27 &  3402 &  0.21 & $-$ \\
$[$ZM00$]$~0028 & 09~59~18.7 & -19~28~22.6 & 15.29 &  3636 &  0.35 & $-$ \\
$[$ZM00$]$~0055 & 09~59~38.5 & -18~55~14.3 & 16.18 &  3661 &  0.69 & Sbc \\
$[$ZM00$]$~0059 & 10~00~37.7 & -19~32~54.7 & 16.30 &  3434 &  0.76 & $-$ \\
$[$ZM00$]$~0069 & 10~00~25.0 & -19~34~59.4 & 16.45 &  3647 &  0.13 & $-$ \\
$[$ZM00$]$~0085 & 09~59~26.5 & -19~38~57.4 & 16.69 &  3538 &  0.34 & Sb  \\
$[$ZM00$]$~0094 & 09~59~44.2 & -19~41~10.9 & 16.78 &  3613 &  0.53 & $-$ \\
$[$ZM00$]$~0136 & 10~00~03.5 & -19~38~24.5 & 17.14 &  3748 &  0.84 & Sc  \\
     Hydra~0030 & 10~00~58.0 & -19~37~44.3 & 19.14 &  3571 &  0.18 & $-$ \\
 $[$C06$]$~2190 & 09~59~28.5 & -19~43~53.0 & 18.56 &  4768 &  0.34 & $-$ \\
 $[$C06$]$~1089 & 09~59~27.1 & -19~45~16.0 & 18.69 &  3614 &  0.27 & $-$ \\
 $[$C06$]$~0694 & 10~00~06.4 & -19~32~49.0 & 18.69 &  5017 &  0.24 & $-$ \\
 $[$C06$]$~2123 & 10~00~39.6 & -19~27~15.0 & 19.75 &  5128 &  0.23 & $-$ \\
 $[$C06$]$~1345 & 09~59~18.6 & -19~24~38.0 & 19.81 &  4509 &  0.23 & $-$ \\
 $[$C06$]$~2234 & 09~59~22.7 & -19~30~29.0 & 20.25 &  3366 &  0.10 & $-$ \\
 $[$C06$]$~0760 & 10~01~05.4 & -19~33~04.0 & 20.27 &  5152 &  0.22 & $-$ \\
 $[$C06$]$~1869 & 09~59~40.8 & -19~27~57.0 & 20.70 &  4846 &  0.23 & $-$ \\
\enddata
\vspace{-20pt}\tablecomments{Top tier:~members, bottom:~associates. 
				ZM00 and C06 list $R$- and $V$-band respectively. 
				Hydra~0030 photometry is in the $R$-band. 
				Morphologies from \citet{rc3}, apart from $[$ZM00$]$~0046 and $[$ZM00$]$~0106,
				characterized using our LCO images. } 
\end{deluxetable}

In any case, HCG~42 shows not only a rich population, but also a far more complex structure than any of the five dwarf galaxy systems we have studied \citep[those in this paper and previous works on HCGs~7 and 59;][]{isk10,isk12a}. The redshifts of sources plotted in Figure~\ref{fig:pspace} are arranged in a continuous distribution, lending support to the interpretation that HCG~42 is a subset of a larger grouping. This might explain the extreme brightness of 42A, which is more akin to that of a bright cluster galaxy (\cf\ the optical brightness of M87). In addition, including all associates in a \mdyn\ calculation increases the mass by five orders of magnitude, which confirms their association not with HCG~42, but the larger collection of galaxies to which HCG~42 itself belongs. The inclusion of members only, however, leads to a physically plausible 20-fold increase. This follows on the previous finding of \citet{rood94} that HCG~42 is associated with the \n3091 group \citep[LGG~186 in][]{garcia93}. It was noted as an `intermediate group' by \citet{decarvalho94}. 

In all, the analysis presented in this section calls for more detailed studies of the extended membership of CGs and similar aggregates. A study of morphological types is essential in this context, to relate groups to clusters and groupings of various densities. In this work morphological typing was only possible for the few galaxies covered in our LCO imaging. Our results might also call for a refinement of the Hickson definition of CGs, given how unstable the routinely employed metrics are to the expansion of group membership.

\section{Summary}\label{sec:summary}
We have presented a multi-wavelength study of the giant and dwarf galaxies that comprise Hickson Compact Groups 16, 22, and 42. Our results can be summarized in three categories: 

\begin{enumerate}

\item[\textbf{1.}]\textbf{Morphological characteristics and indications of past dynamical events}. Starting with an examination of large-scale morphology, we found a number of noteworthy traits among the eleven giant galaxies. HCG~16B is lopsided, perhaps as a result of a recent interaction with 16A. Galaxies 16C and 16D present irregular morphologies from the ultraviolet to the infrared. HCG~16D is known to exhibit X-ray emission consistent with the \ha\ emission detected by \citet{werk10} in the form of a super-galactic wind \citep{rich10,vogt13}. The `secondary nuclei' reported by \citet{decarvalho99} are likely to be agglomerates of star clusters whose light was blended by the $1\farcs5$ seeing. 
 
Within the large H\one\ envelope of HCG~16 we also find a tidal tail, extending to the east of 16A. While our interpretation needs to be confirmed with deeper imaging, this feature presents a good opportunity to study a debris feature at the end of its optically detectable phase. We find a hint of ongoing star formation at the tip of the tail, while the rest of the feature appears to host a $\lesssim1~$Gyr-old stellar population. The existing data do not allow us to discern between different evolutionary scenarios --  old stars stripped in an interaction versus in-situ star formation a long time ago. 

All three galaxies in HCG~22 display interesting traits: the equatorial dust ring in 22A and the network of low surface brightness features around 22B are consistent with recent merger or infall events, while the bright central bar and extremely faint, loose network of spiral arms of 22C represent the only sites of star formation in the system -- and also where all the H\one\ is situated. The galaxies of HCG~42 do not display any diversions from quiescence.

\item[\textbf{2.}]\textbf{Star cluster populations and star formation histories}. The evolutionary sequence mapped out by the three CGs in terms of gas richness is reflected in the star clusters. The color distribution of the cluster populations portray a `young' HCG~16 that has yet to develop a large globular cluster population, as opposed to `old' HCG~42, where the bimodal GC population accounts for the majority of detected sources. There is a distinct lack of young clusters here ($<1~$Gyr), which temporally marks the quenching of star formation in HCG~42. In between the two groups is HCG~22, which hosts a very unusual stellar population, dominated by clusters of intermediate age. This reveals that the bulk of star formation activity happened over the past few Gyr, with little recent activity, practically none outside the spiral arms of 22C. To our knowledge, no individual galaxy or grouping has been found to host such a markedly intermediate-age population. This is an observer bias introduced by the preferential study of the young cluster populations of either mergers and highly star forming systems, or the GC populations of early-type galaxies. From their morphologies, we propose that galaxies 22A and 22B are recent merger remnants. The star cluster populations are consistent with this scenario. 

The above provides a good demonstration of the utility of star clusters as chronometers for past star formation events. Any epoch at which a galaxy exhibits a significant star formation rate will be recorded by the star clusters, and bursts will register as `bumps' along the model track \citep[see the analysis of Stephan's Quintet by][]{fedotov11}. Equally important is the dearth or absence of clusters in certain parts of color space, indicating a low or zero SFR, such as HCG~42. This provides an estimate of the time when a system entered a mode of galaxy evolution where no more gas is being converted to stars. At this stage, any gas in the IGM can only be redistributed or heated. 

\item[\textbf{3.}]\textbf{Dwarf galaxy membership and implications}. The information on dwarf galaxies collected in this study originates from a variety of sources, however, all utilized catalogues are complete to $R\simeq18~$mag, or $M_R=[-15.5, -14.7, -15.9]~$mag. This gives us the opportunity to study the bright end of the dwarf galaxy luminosity function and place the three groups in the context of the distinction between isolated and embedded groups \citep{mcconnachie09cg,mendel11}. HCGs~16 displays a more or less symmetric spatial distribution of dwarfs, albeit all but one (HCG~16X) lie outside the H\one\ envelope shared by the giants. This might indicate that the gas was released through interactions between the giants. The phase-space diagram of HCG~42 places it at the end of a filament or other complex velocity structure made of $\approx50$ dwarf galaxies. 

We made two cuts in velocity space, at $3\,\sigma$ and $10\,\sigma$, to sort between members and `associates'. This way we were able to test the applicability of the velocity dispersion, designed to characterize populous galaxy clusters, in studying small groups. In other words, what is the meaning of a velocity dispersion derived from only four galaxies? This question is especially relevant seeing as CGs are not necessarily expected to be dynamically relaxed. 

Indeed, the inclusion of all `associate' dwarf galaxy velocities gives rise to a tremendous change in the inferred dynamical properties of the three groups. The dynamical masses (\mdyn, which are only truly appropriate for virialized systems) increase by factors of $4$, $50$, and a gargantuan $10^5$ for HCGS 16, 22, and 42. In contrast, including only high-confidence members (the $3\,\sigma$ cut) only increases the \mdyn\ by a factor of $\sim10$ for all three groups. This indicates that, although flawed from a statistical perspective, the velocity dispersion of the group core might provide a reliable metric. More groups need to be studied in order to ascertain this eventuality.

Another way to examine this effect of dwarfs on the \mdyn\ derivation is in terms of the hierarchy in which a group is found. While the isolated HCGs~16 and 22 are mildly affected by the inclusion of dwarfs \textit{and} associates, the derivation of \mdyn\ breaks down for embedded HCG~42. In all, we find that updating the velocity dispersion through the careful inclusion of high-confidence dwarf members vastly upgrades its value by making it a statistically viable metric. 

\end{enumerate}

\acknowledgements We thank the anonymous referee for the enthusiastic reception of our work and for suggestions that bettered the manuscript. Funding was provided at PSU by the National Science Foundation under award AST-0908984. Support for this work was provided by NASA through grant number HST-GO-10787.15-A from the Space Telescope Science Institute which is operated by AURA, Inc., under NASA contract NAS 5-26555. 
KF and SCG thank the Natural Science and Engineering Research Council of Canada and the Ontario Early Researcher Award Program for support.
This paper makes use of publicly available SDSS imaging and spectroscopy. Funding for the creation and distribution of the SDSS Archive has been provided by the Alfred P. Sloan Foundation, the Participating Institutions, the National Aeronautics and Space Administration, the National Science Foundation, the U.S. Department of Energy, the Japanese Monbukagakusho, and the Max Planck Society. The SDSS Web site is http://www.sdss.org/. The SDSS is managed by the Astrophysical Research Consortium (ARC) for the Participating Institutions. The Participating Institutions are The University of Chicago, Fermilab, the Institute for Advanced Study, the Japan Participation Group, The Johns Hopkins University, Los Alamos National Laboratory, the Max-Planck-Institute for Astronomy (MPIA), the Max-Planck-Institute for Astrophysics (MPA), New Mexico State University, Princeton University, the United States Naval Observatory, and the University of Washington.
This research has made use of the NASA/IPAC Extragalactic Database (NED) which is operated by the Jet Propulsion Laboratory, California Institute of Technology, under contract with the National Aeronautics and Space Administration. 
\bibliographystyle{apj}
\bibliography{references}

\end{document}